\documentclass{JHEP3}

\usepackage{latexsym,amsfonts,amssymb,amsmath,amsthm,graphicx,array}

\newcounter{fignum}
\renewcommand{\caption}[1]{\def\captiontext{#1}}
\renewenvironment{figure}{
\def\captiontext{}
\refstepcounter{fignum}
\par\noindent
}{
\nopagebreak
{\centering\par}
\noindent
Fig.~\thefignum:~\captiontext
\par\noindent
}
\def\beq{\begin{equation}}                     %
\def\eeq{\end{equation}}                       %
\def\bea{\begin{eqnarray}}                     %         %
\def\eea{\end{eqnarray}}                       %       %
\def\PL{{\it Phys. Lett.} }                    %         %
\def\PR{{\it Phys. Rev.} }
\def\PRL{{\it Phys. Rev. Lett.} }
\def\APJ{{\it Astroph.~J.~}}
\def\AJ{{\it Astron.~J.~}}                 %

\DeclareMathOperator{\sech}{sech}

\newcommand{\pd}{\partial}
\def\beq{\begin{equation}}                     %
\def\eeq{\end{equation}}                       %
\def\bea{\begin{eqnarray}}                     %         %
\def\eea{\end{eqnarray}}

\title{Time Lumps in  Nonlocal Stringy Models and Cosmological Applications}

\author{
I.~Ya.~Aref'eva\footnote{\texttt{arefeva@mi.ras.ru}, Steklov
Mathematical Institute, Russian Academy of Sciences}~~and~
L.~V.~Joukovskaya\footnote{\texttt{ljoukov@mi.ras.ru}, Steklov
Mathematical Institute, Russian Academy of Sciences and V\"axj\"o University,
Sweden.}}

\abstract{
We study lump solutions in nonlocal toy  models
and  their cosmological applications.
These models are  motivated by a
 description of  D-brane decay within string field theory framework.
In order  to find
cosmological solutions we use the simplest local approximation
keeping only  second derivative terms in nonlocal dynamics.
 We study validity of this approximation in  flat background
 where time lump solutions can be written explicitly and  work out the validity of this
approximation.
Finally, we show that  our models at large
time
exhibit the
phantom behaviour similar to the case of the string  kink.
}

\begin{document}

\tableofcontents

\section{Introduction}
%\label{sec:int}
%\setcounter{equation}{0}
In the last years strings and D-branes attracted cosmological
applications related with the cosmological acceleration
\cite{Sen}-\cite{IA1}. The observations suggest that the Universe
is presently accelerating \cite{Perlm,Riess}. It is believed that
a new physics is required to explain this phenomenon. The bulk of energy density in the
Universe is gravitationally repulsive and can be represented as
 an unknown form of energy (dark energy) with negative pressure and

negative equation of state parameter $w=p/\rho$. The current
experimental data show that $w$ lies in the range $-1.61<w<-0.78$
\cite{knop,Spergel,Tegmark}. There are various theoretical models
-- a quintessence scalar field \cite{Sachni},
Dirac-Born-Infeld action \cite{Sen}-\cite{FS}, and other  which are able to
describe the case $w>-1$.

The most exciting  possibility would be the case $w < -1$ (see
\cite{Sachni}-\cite{AKV},\cite{IA1}
 and references therein). There are several phenomenological
models describing this phantom Universe
\cite{Caldwell},\cite{Caldwell03}. In these models week energy
conditions $\rho>0, \rho+p>0$ are violated and most of them are
unstable and  strange phenomena  related with unstability appear
\cite{carroll03}. There are also  models related with modified
gravity \cite{mod-grav}, but there is a problem of getting these
models basing on  fundamental principles.

Recently it was shown  that the  string field theory (SFT)
description of D-brane decay leads at large times to an effective
phantom model \cite{IA1}. As a result of this decay the spectrum
becomes stable and the model does not suffer from instability
problems. D-brane decay  in non-flat metric within  the SFT
framework is described by the string tachyon action \cite{IA1}
\begin{equation}
\label{action-IA} S = \int \sqrt{-g}d^4x\,\left(
\frac{\alpha'}{16\pi G}R+\frac{1}{g_0^2}(-\frac{\kappa^2}{2}
g^{\mu\nu}\partial_\mu\phi\partial_\nu\phi+ \frac12 \phi^2-U(\Phi)
)\right),
\end{equation}
Here it is supposed that we deal with 3-dimensional D-brane, $G$ is
a $4$--dimensional gravitational constant, $\alpha '$ is a
string tension, $g_{\mu\nu}$ is a metric for dimensionless
coordinates,
 $R$ is a curvature,  $g_0$ is a
dimensionless string coupling constant, fields $\phi$ and $\Phi$
are dimensionless fields related via
\begin{equation}
\label{met} \Phi=\exp(\frac {\beta }{2}\Box_g) \phi,~~~~~~
\Box_g=\frac{1}{\sqrt{-g}}\pd _\mu \sqrt{-g}g^{\mu\nu} \pd _\nu,
\end{equation}
$\beta$ and $\kappa^2$ are  constants whose values depend on the
type of a string  under consideration. The form of the potential
also depends on the type of a string.

Action (\ref{action-IA}) in the Friedmann background
\begin{equation}
\label{F-m} ds^2=-dt^2+a^2(t)(dx^2_1+dx^2_2+dx^2_3),
\end{equation}
gives  Friedman equations and  the equation for the field
 $\Phi=\Phi(t)$ which in turn reads as
 \begin{eqnarray}
\label{Psi-ap}
 \left(\kappa^2{\cal D}+1\right)e^{-\beta {\cal D}}\Phi =U'(\Phi),
\end{eqnarray}
where $ {\cal D}=-\partial _t^2-3H(t)\partial_t $, $H=\dot{a}/a$
is a Hubble parameter  and $U'(\Phi)=\partial U/\partial \Phi$.

To study solutions of equation (\ref{Psi-ap})  the following
approximation was used \bea \label{e.o.m.} (-\kappa ^2+\beta
\kappa_{0}
^2)(\ddot{\Phi}+3H\dot{\Phi})=V'(\Phi),\\
\label{pot} V(\Phi)=U(\Phi)-\frac{1}{2}\Phi^2, \eea $\kappa^2
_{0}$ is a parameter close to 1.  This
 second
order derivatives approximation is  called mechanical or local
approximation. In a rather rough approximation one can take
$\kappa^2 _{0}\simeq 1$ (a so-called  direct mechanical
approximation).

For  $\beta\kappa^2_{0}>\kappa ^2$ we get ghost kinetic term in
(\ref{e.o.m.}). For this case after time rescaling we can take
 $\beta\kappa^2_{0}-\kappa ^2=1$. This case corresponds to a
phantom model. The dynamics of the  phantom model in the flat
background reproduces dynamics of the usual particle in the
overturn potential \cite{AJK}. The equation of  state parameter
$w$ for  the phantom in a general potential can be represented as
\bea \label{omega-H}
w=-1-\frac{1}{3m^2_p}\frac{\dot{\Phi}^2}{H^2}, \eea where $m_p^2$
is a dimensionless parameter related with the Plank mass $M_p^2$,
$m_p^2=M_p^2\alpha '=\alpha '/8\pi G$.

For a cubic potential $U$ the action (\ref{action-IA}) is the
curved version of the action for the tachyon field in the Witten
SFT \cite{Witten-SFT} at the lowest order in  the level truncation
scheme \cite{Kost-Sam,West}.  It is supposed  that we deal with
3-dimensional D-brane in 26-dimensional space-time
 and the volume
of compactified 22-dimensional space in (\ref{action-IA}) is
omitted.

For a quartic $U$ the action (\ref{action-IA}) is the curved
version of the action for the tachyon field which was obtained in
an approximation of slow varying auxiliary field \cite{AJK} from
cubic fermionic SFT\cite{AMZ-PTY}.  This action  in the flat case has been obtained for
the case when the string field contains only two fields: the
tachyon in the GSO(-) sector\footnote{To produce
nonBPS-branes  the GSO(-) sector is involved
\cite{BSZ},\cite{ABKM} (see reviews \cite{0102085}-\cite{0301094}
for details).} and a lowest auxiliary field in the
GSO(+) sector. Integration over this auxiliary field produces a
fourth degree  of the tachyon self interaction.

The existence of the kink solution for the flat version of
equation (\ref{Psi-ap}) with $\kappa=0$ has been  shown
numerically in \cite{BFOW} and has been proved in \cite{VS-YV}. A
validity of the approximation (\ref{e.o.m.}) for this kink
solution has been studied in \cite{Yar}.

In the case of the flat metric and $\kappa= 0$ the kink solution
of (\ref{Psi-ap}) is a monotonic function interpolating between
two different vacuum solutions. However not only monotonic kink
solutions appear in the dynamics of string tachyons. For
$\kappa\neq 0$, $\kappa^2<\beta \kappa^2 _{0}$ and  $H=0$ there are
non-monotonic kink solutions of  (\ref{Psi-ap}) and these
solutions have bounce points. In the open-closed string model
\cite{OH}
 a bounce point  also
appears \cite{LY}. Typical examples of solutions with one bounce
point are lump solutions. In particular, such type of solutions has been recently found in
\cite{FGN}.

In this paper we consider two  SFT  inspired models with the
action
 (\ref{action-IA}). For the first model
 \begin{equation}
\label{UI} U=\frac{\sqrt{k}}{k+1}\Phi^{k+1}, ~~~~~k<1,
\end{equation}
 and for the second  model the potential is a special polynomial
 \begin{equation}
\label{UII} U=\sum_{n=1}^{N}\frac{\alpha_n}{n+1}\Phi^{n+1}
\end{equation}
Distinguished feature of these models is that they have explicit
lump solutions for $\kappa=0$ in  flat background. We study a
validity of   the mechanical approximation to these solutions. For
this purpose we compare  essential physical characteristics such
as energy and pressure of the nonlocal systems with their
mechanical analogues. We find that under special conditions these
characteristics in both models are rather similar. Also in both
cases it is possible to find
 small explicit modifications of the mechanical potentials to
 solve  corresponding local equations  in the Friedmann metric.

The paper is organized as follows. In Section 2 we consider the
equation of motion
\begin{equation}
\label{eqt} e^{\beta \partial ^2_t}\Phi=\sqrt{k}\Phi^k.
\end{equation}
where $0<k<1$, $a>0$. There is the lump solution to this equation
\begin{equation}
\label{sol-I} \Phi(t)=e^{-\xi t^2},~~\xi=\frac{1-k}{4\beta k},
\end{equation}
The energy  of the solution (\ref{sol-I})  is equal to zero.

In Section 2.2 we compare (\ref{sol-I}) with zero energy periodic
trajectories in
 the direct mechanical approximation
of (\ref{e.o.m.}). The period $T_k$ becomes large when $k\to 1$.
We show that for time $t < T_k/2$ the zero-energy  periodic
solutions to the mechanical problem rather well approximate  the
lump solution (\ref{sol-I}).

In Section 2.3 we calculate explicitly   the pressure on the
solution (\ref{sol-I}) and in Section 2.4 compare it with the
pressure of the  zero-energy periodic solution of the mechanical
problem for $|t|<T_k$. We find that the mechanical approximation
represents
 a behavior of the pressure  of
 the non local problem with a good degree of accuracy
 for $k$ closed to 1, $k<1$.
We also compare $\Phi$ with $\tilde{\Phi}=e^{\beta /2\partial
^2_t}\Phi$
 and  conclude that  a comparison of $\Phi$ and
$\tilde{\Phi}$  gives  a rather good quantitative criterium for a
validity
 of the local approximation. A precise condition
of a validity of the local approximation is a condition that the
magnitude of $\tilde{\Phi}-\Phi-\frac{\beta}{2}\partial_t ^2\Phi $ is
small.

In Section 3.1 we consider   a modified integral equation
 with the constant friction
\begin{equation}
\label{eqt}
e^{\beta  \partial ^2_t+h\partial
_t}\Phi_h=\sqrt{k}g(t)\Phi_h^k.
\end{equation}
and special time depending coupling constant  $g(t)=e^{-2h(1-k)t}$.
(\ref{eqt}) has an  explicit lump solution
\begin{equation}
\label{sol-If} \Phi_h(t)=e^{-\xi t^2+Bt},~~B=\frac{h}{2\beta
}(-1-\sqrt{1+16\xi \beta ^2}),
\end{equation}
We see that the friction makes behavior of the solutions of the
nonlocal equation more closed to that of the
 corresponding mechanical problem.
 This give us a possibility to use this approximation for Friedmann equations.

In Section 3.2 we study numerically the   corresponding Friedmann
equations and find that there is a regime, i.e. suitable initial
conditions, that represents an acceleration with $w<-1$.

In Section 3.3 we reconstruct explicitly the potential for which
(\ref{sol-I}) is  an exact solution of the Friedman equations.
This potential has the form
\begin{equation}
\label{pot-f} V(\Phi)=V_0(\Phi)+V_1(\Phi),
\end{equation}
where
\begin{equation}
\label{pot-f0} V_0(\Phi)=-2\,\xi \,{\Phi }^2\,\log \Phi,
\end{equation}
and  $\delta V (\Phi)$ is of order of $1/m^2_p$.
$$V_1 (\Phi)=\frac{3\,\xi }{64 m_p^2}\left(\sqrt{2\pi }\,
          {\mbox e~rf}(\sqrt{-2\log \Phi }) -
         4\Phi ^2\,\sqrt{-\log \Phi } \right)^2.$$

In Section 4 we search for lump solutions of the following
equation
\begin{equation}
\label{eqII}
 e^{\beta  \partial ^2}\Phi=\sum_{n=1}^{N}\alpha_n \Phi^n
\end{equation}
where $\alpha_n$ are  constant. It is know that a direct numerical
search for lumps solutions is a rather difficult problem.  There
are also nonexistence theorems for a wide class of
potentials\cite{MZ,VS-YV}. It is also interesting to find lump as
well as kink solutions in vacuum SFT \cite{LB,HM}.

Here we accept the following strategy. We start from a given
function
\begin{equation}
\label{solII} \varphi_{0}(t)=\sech^2(t)
\end{equation}
and find constant $\alpha_n$ from a requirement that the function
(\ref{solII}) is a solution to equation  (\ref{eqII}) with a small
discrepancy for fixed $N$. We mean that the discrepancy has a
small $L_2$-norm.   We  consider the cases $N=3,...,14$.
Qualitatively one can say that lowest coefficients  $\alpha_n$
show a stable picture when $N\to \infty$.
 With the obtained $\alpha_n$ in Section 4.2 we find
numerical solutions to (\ref{eqII}) with zero energy using an
analog of Freedman method (see \cite{LY} and references therein).

Then in Section 4.3 we  study a  mechanical approximation to
equation (\ref{eqII}). It  occurs that all  potentials $V$ have
local maximum at $\Phi=0$ and  are equal to zero at $\Phi_{0N}$,
$\Phi_{0N}<1$ and $\Phi_{0N-1}\simeq \Phi_{0N}$. As has been
mentioned above, the lowest coefficients $\alpha_n$  are
stabilized as $N \to\infty$. For the zero-energy solutions of the
mechanical problem with the initial data $\Phi(0)=\Phi_{0N}$ one
has  $|\Phi(t)|<1$ and for such $\Phi$ only the low-order
coefficients dominate in the potential. Therefore, in spite of а
global change of the behavior of potentials  with N increasing,
the form of potentials near the region, where the particle with
the above initial data moves, does not change. For small time
trajectories of the mechanical problems with above mentioned
initial data do not reproduce
 the lump (\ref{solII})
since $\sech^2(t)$  has derivatives of the same order as itself.
However
 for large time when trajectories are closed to  an attractor point
 $\Phi=0$ they are also closed to the lump.

The profile of the potential (\ref{pot}) with found $\alpha_n$
 in the region $0<\Phi<1$
is rather closed to the profile of the potential
\begin{equation}
\label{pot-sech} V_0(\Phi)=2\,\left( 1 - \Phi  \right) \,{\Phi }^2
\end{equation}
for which the function (\ref{solII}) solves the mechanical problem
with the lump boundary conditions. In Section 4.5 we show that
engaging of  the Friedmann metric modifies the potential
(\ref{pot-sech}) and one has  to add
 $$V_1=-
  \frac{4}{75m_p^2}{\left( -1 + \Phi  \right) }^3\,
     ( 2 + 3\Phi)^2$$
to (\ref{pot-sech}) in order the function (\ref{solII})
 to  be a solution of the  Friedmann equations.

%%%%%%%%%%%%%%%%%%%%%%%%%%%%%%%%%%%%%%%%%%%%%%%%%%%%%%%%%%%%%%%%%
\section{Gaussian Lump}
\label{sec:exp} \setcounter{equation}{0}
\subsection{Action and equation of motion}
In this section we consider the action
\begin{equation}
\label{action-exsol} S = \int d^d x \left[\frac{1}{2}\phi^2-
\frac{\sqrt{k}}{k+1}\Phi^{k+1}\right],
\end{equation}
with an unusual non-polynomial interaction
\begin{equation}
\label{k} 0<k<1
\end{equation}
and where $~\Box= -\partial^2 +\nabla^2,$
\begin{equation}
\label{Pp} \Phi(x)=e^{\frac{\beta }{2}\Box} \phi(x), ~~~a > 0,
\end{equation}
 The equation of motion for this action is
\begin{equation}
\label{eq1} e^{-\beta  \Box}\Phi(x)=\sqrt{k}\Phi^k.
\end{equation}
For space homogeneous configurations $\Phi=\Phi(t)$ it takes the
form
\begin{equation}
\label{exeq-h} e^{\beta
\partial^2}\Phi(t)=\sqrt{k}\Phi^k(t),\end{equation}
 ~here~ and~ below~
$\partial=\partial_t$.

Equation (\ref{eq1}) has the following solution
\begin{equation}
\label{P-sol}
 \Phi(t)=e^{-\frac{b}{4\beta }t^2},~~b=\frac{1-k}{k}
\end{equation}
The relation between $\Phi$ and $\phi$ is
\begin{equation}
\Phi(t)=e^{-\frac{\beta }{2}\partial^2 } \phi(t)
\end{equation}
and  one can get $\phi$ from $\Phi$ using the smoothing integral
representation
\begin{equation}
\label{Pp} \phi(t)=\frac{1}{\sqrt{2\pi}}\int
e^{-\frac{(t-t')^2}{2\beta }}\Phi(t')dt'
\end{equation}
Substituting (\ref{P-sol}) into (\ref{Pp}) one gets smoothed
solution $\phi$
\begin{equation}
\label{p-sol}
\phi(t)=\frac{\sqrt{2}}{\sqrt{2+b}}e^{-\frac{b}{2\beta (2+b)}t^2}.
\end{equation}

\begin{figure}
\includegraphics[width=6.5cm]{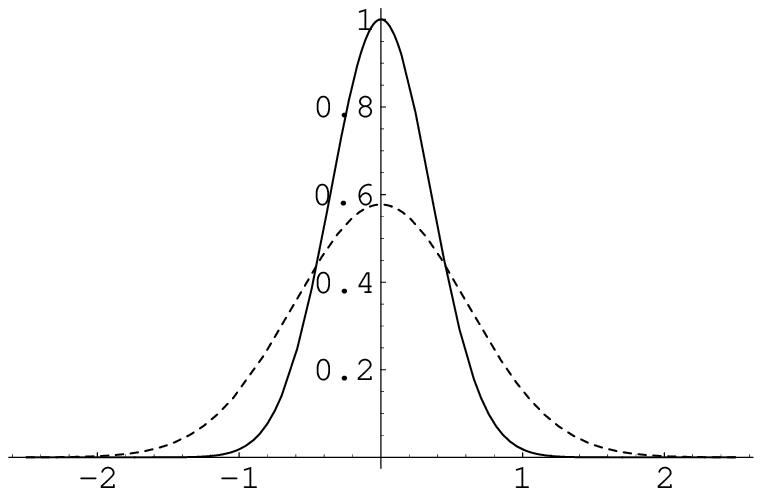}
1)~~
\includegraphics[width=6.5cm]{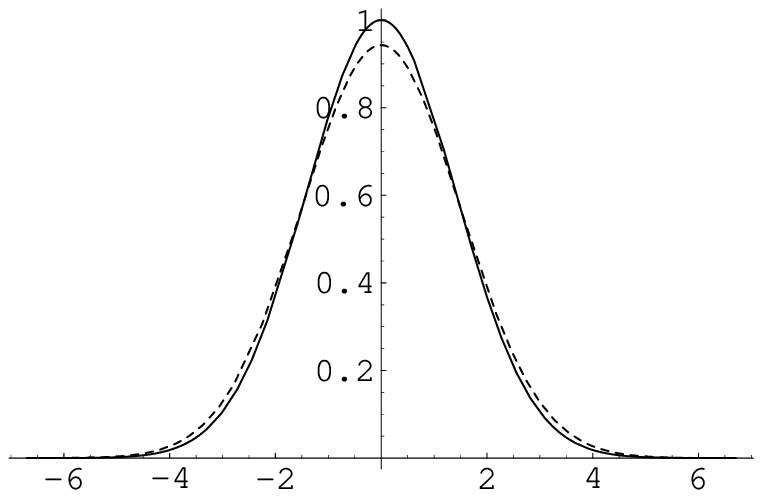}
2)~~
\caption{
\label{fig:ExpKExp.eps}
Solutions of integral equation (\ref{exeq-h}):
1) for $\beta=\frac{1}{4},$  $k=0.2$;
2) for $\beta =\frac{1}{4},$  $k=0.8$.
The firm line is  $\Phi$ and the dashing line is $\phi$.
}
\end{figure}

From (\ref{P-sol}) and  (\ref{p-sol}) we see that $\Phi$ and $\phi$ are related as
\begin{equation}
\label{pP} \phi(t)=\sqrt{c}\Phi(\sqrt{c}
t),~~~\mbox{where}~c=\frac{2k}{1+k}
\end{equation}
 When
$k\to 1$, $c\to 1$ and therefore $\phi \cong \Phi$ \footnote{Let
us note that
 in the more complicated model \cite{OH} ($c_{2}=\frac{13}{6}$)
it is  known, that physical (analog of $\phi$) and tilded (analog
of $\Phi$) fields coincide with very high precision \cite{LY}.}
and when $k \to 0$, $c\to 0$ and $\phi \cong 0$. This is
illustrated on Fig. \ref{fig:ExpKExp.eps} where solutions of
(\ref{exeq-h}) and corresponding smoothed fields $\phi$ are
presented for different values of $k$.

\subsection{Mechanical approximation}
\label{eq-t1-exsolmechpr} 
\begin{figure}
\includegraphics[width=4.3cm]{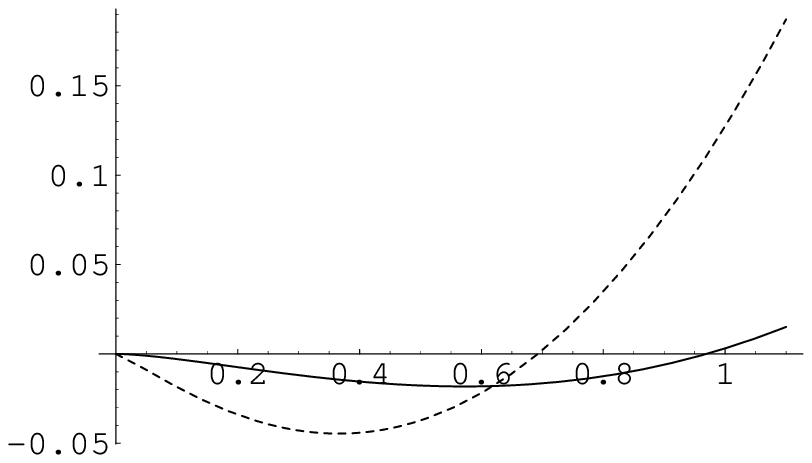}a)
\includegraphics[width=4.3cm]{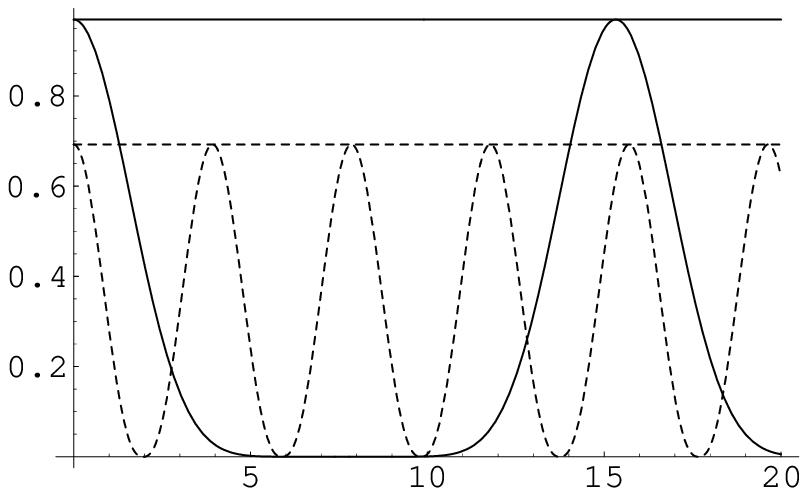}b)
\includegraphics[width=4.3cm]{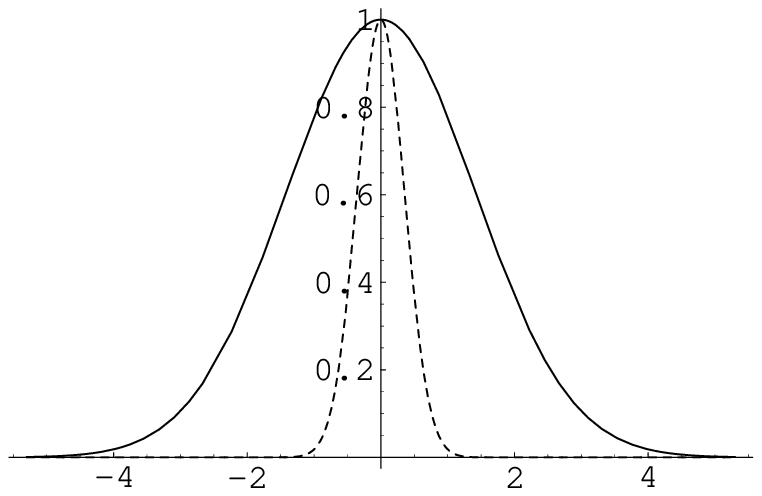}~c)
\caption{a) The  potential $V$ for the  k= 0.2 (dashing line) and
for k=0.8 (firm line). b) The numerical solution of equation
(\ref{eq1-mech}) for k=0.2 (dashing line) and for $k=0.8$ (firm
line). c) Exact solution to integral equation (\ref{exeq-h}) for
$\beta=1/4$ and k= 0.2 (dashing line) and  k=0.8 (firm line).
\label{fig:potlump}
}
\end{figure}

Let us consider the mechanical
approximation for equation (\ref{exeq-h}),
\begin{equation}
\label{eq1-mech} \beta\ddot{ \varphi}(t)=-V'_{ot}(\varphi),
\end{equation}
the parameter $\beta $ plays a role of  mass of the particle. The
potential $V_{ot}$  is an overturn version of the potential
(\ref{pot}) for $U$ given by (\ref{UI}),
\begin{equation}
\label{effpot1} V_{ot}(\varphi)=-V(\varphi), ~~~V(\varphi)
=\frac{\sqrt{k}}{k+1}\varphi^{k+1}-\frac{1}{2}\varphi^2,
\end{equation}
To have a real potential $V$ in (\ref{effpot1}) let us assume that
$k=1-2\delta$ and present the corresponding square in $U$ as
$(\Phi ^2)1/(1-\delta)$. In this case the potential $V_{ot}(\Phi
)$ is an even function and this function is equal to zero at $\Phi
=0$ and has а beak with  singular second derivative $\Phi =0$, see
Fig.\ref{fig:potlump}.a. The particle with the zero energy
oscillates between the  point $ \varphi=0$ and the bounce at
$\varphi_{max}=(\frac{2\sqrt{k}}{1+k})^{\frac{1}{1-k}}$. The
zero-energy  trajectories
 are presented in the
fig.\ref{fig:potlump}.b and the exact lumps (\ref{P-sol})  are
presented on fig.\ref{fig:potlump}.c for different $k$.

The periods of these motions are
\begin{equation}
T_k=\sqrt{\beta }\int_{0}^{\varphi_{max}} \frac{d\varphi}
{\sqrt{-2V_{ot}}}.
\end{equation}
In the table below we enumerate  values of $T$ for different  $k$
and $\beta =1/4$.

\begin{center}
\begin{tabular}{|c|c|c|c|c|c|c|}
\hline& & & & & &\\
$k$ & 1/5& 1/3 & 1/2 & 2/3& 8/9
 & 9/10\\
\hline& & & & & &\\
$T$ & 5$\pi$/4 & 3$\pi/2$ & $2\pi$ & $3\pi$ & $9\pi$ & $10\pi$\\
\hline
\end{tabular}
\end{center}

The solutions of the integral equations (\ref{exeq-h})(see Fig.
\ref{fig:potlump}.c)  and their mechanical analogs
(\ref{eq1-mech}) describe different physical behavior. Namely, the
solution (\ref{P-sol}) of the integral equation starts from zero
go above the bounce point $\varphi_{max}$ then rises $\Phi=1$ and
returns to zero. Thus the solution of the integral equation does
not bounded by the bounce point, whereas the zero-energy solution
of the mechanical model presents a  periodic motion between the
bounce point and zero. The period depends on $k$ and as $k$ goes
to $1$ the period becomes infinite, i.e.  $k=1$ corresponds to the
free motion. Therefore, for small $k<<1$ the mechanical
approximation does not work, while it works for $t<T_k$ for $k$
goes to 1.

\subsection{Energy  and pressure}
In this subsection we consider the energy and the pressure of the
nonlocal problem (\ref{exeq-h}) and compare them with energy and
pressure of the corresponding  mechanical problem
(\ref{eq1-mech}). Equation (\ref{eq1}) has the conserved energy
(compare with \cite{AJK})
\begin{equation}
\label{E} E=E_{p}+E_{nl_1}+E_{nl_2},
\end{equation}
where
\begin{equation}
\label{Ep1}
E_{p}=-\frac{1}{2}\phi^2+\frac{\sqrt{k}}{k+1}\Phi^{k+1},
\end{equation}
\begin{equation}
\label{En1} E_{nl_1}=\frac{\beta }{2}\int_{0}^{1}(e^{-\frac{\beta
}{2} \rho
\partial^2} \sqrt{k}\Phi^{k})
 (e^{\frac{\beta }{2} \rho \partial^2} \partial^2 \Phi) d \rho
\end{equation}
and
\begin{equation} \label{En2}
E_{nl_2}=-\frac{\beta }{2}\int_{0}^{1}((e^{-\frac{\beta }{2} \rho
\partial^2} \sqrt{k}\partial \Phi^{k})) (e^{\frac{\beta }{2} \rho
\partial^2} \partial \Phi) d \rho.
\end{equation}
One can see that this energy conserves. Indeed,
$$
\frac{dE(t)}{dt}=-\phi \partial \phi+ \sqrt{k}\Phi^k \partial \Phi
+\frac{\beta }{2}\int_{0}^{1}(e^{-\frac{\beta }{2} \rho
\partial^2} \sqrt{k}\Phi^{k}) \overleftrightarrow{\partial}
(e^{\frac{\beta }{2} \rho \partial^2} \partial \Phi) d \rho$$
$$
=-\phi \partial \phi+ \sqrt{k}\Phi^k \partial \Phi -
\sqrt{k}\Phi^{k}\overleftrightarrow{e^{-\frac{\beta }{2}
\partial^2}}
\partial \phi =\partial \phi\left[-\phi+
 e^{-\frac{\beta }{2} \partial^2}\sqrt{k} \Phi^{k}\right]=0.
$$

 The
pressure in the model (\ref{action-exsol}) has the form
\begin{equation}
\label{P1} P=E_{nl_2}-E_p-E_{nl_1}
\end{equation}
Substituting in this formula expressions for $E_{nl_1}$ and
$E_{nl_2}$ given by (\ref{En1}) and (\ref{En2}) one gets
\begin{equation}
\label{pressure-twoparmeter} P=-\beta
\int_{0}^{1}(e^{(2-\rho)\frac{\beta }{2}  \partial^2}\partial
\Phi) (e^{\frac{\beta }{2} \rho \partial^2} \partial \Phi) d \rho
\end{equation}
Substituting  in (\ref{pressure-twoparmeter}) the explicit form of
the solution (\ref{P-sol}) we get
\begin{equation}
\label{pressure-int} P=-\frac{2  b^2t^2}{\beta } \int_{0}^{1}
\frac{d \rho } {[(2-b(\rho-2))(2+b \rho)]^{\frac{3}{2}}}~
\exp(-\frac{t^2}{\beta }\frac{b(2+b)}{4+4b-b^2(\rho-2)\rho}) ,
\end{equation}

From formula (\ref{pressure-int}) we  see that the pressure is
negative and for large time goes to zero.
% (see fig.\ref{fig:exactpressure.eps}).

\subsection{Compression of   pressure for nonlocal and local problems}
Comparing expressions (\ref{E}) and (\ref{P1}) with  standard
expressions of the energy and pressure for a scalar field,
$E_{\varphi}=\frac12 \dot{\varphi}^2+V(\varphi)$,
$P_{\varphi}=\frac12 \dot{\varphi}^2-V(\varphi)$, one can say that
$E_{nl_1}$ and $E_{nl_2}$ play roles of extra terms in the
nonlocal "kinetic' energy  ${\cal E}_k $ and 'potential' energy
${\cal E}_p$, i.e.
\begin{equation}
\label{Ec} E={\cal E}_k+{\cal E}_p
\end{equation}
\begin{equation}
\label{pc} P={\cal E}_k-{\cal E}_p
\end{equation}
and
\begin{equation}
\label{Ek} {\cal E}_k=E_{nl_2}
\end{equation}
\begin{equation}
\label{Ep} {\cal E}_p=E_p+E_{nl_1}
\end{equation}
where $E_p$ and $E_{nl_1}$, $E_{nl_2}$ are given by (\ref{Ep1}-\ref{En2}).

\begin{figure}
\includegraphics[width=4.9cm]{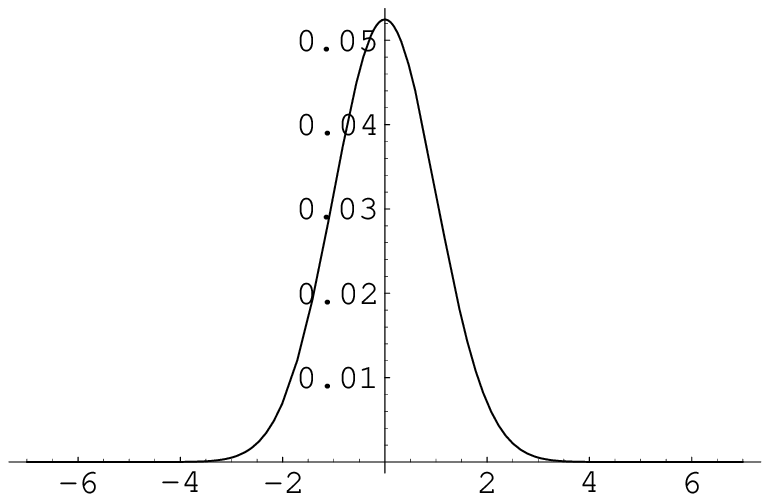}~1a~~
\includegraphics[width=4.9cm]{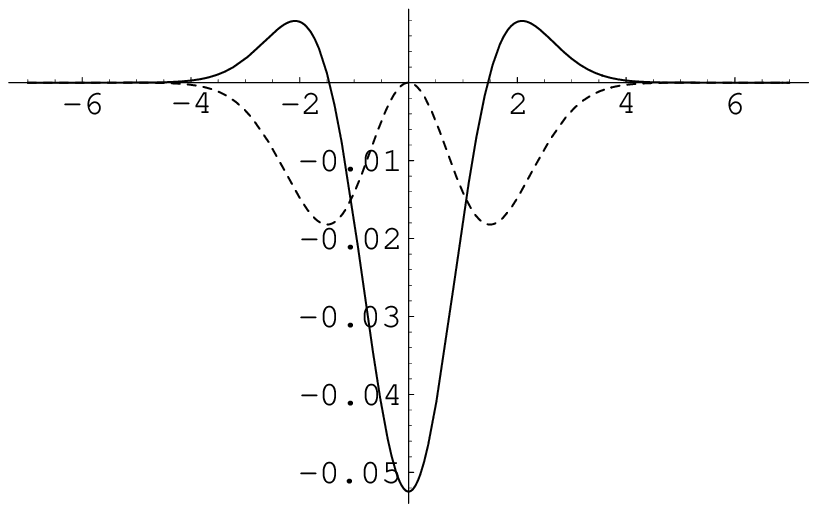}~1b~~
\includegraphics[width=4.9cm]{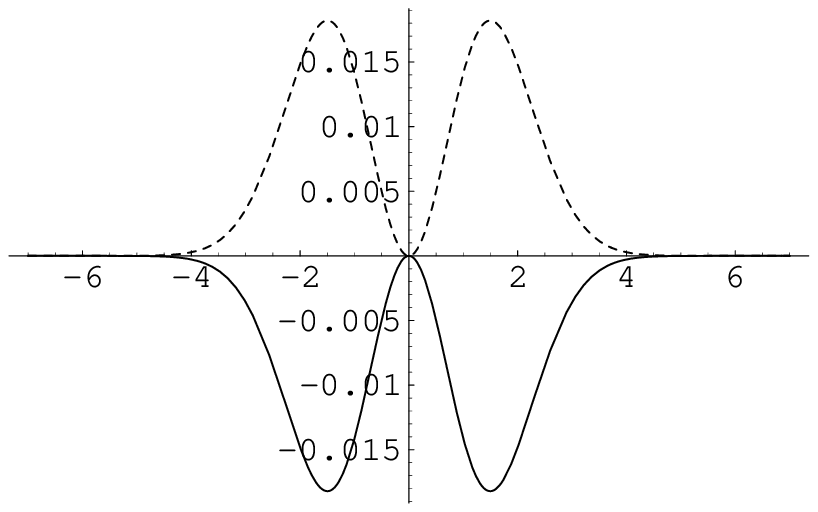}~1c~~\\
$~$\\
\includegraphics[width=4.9cm]{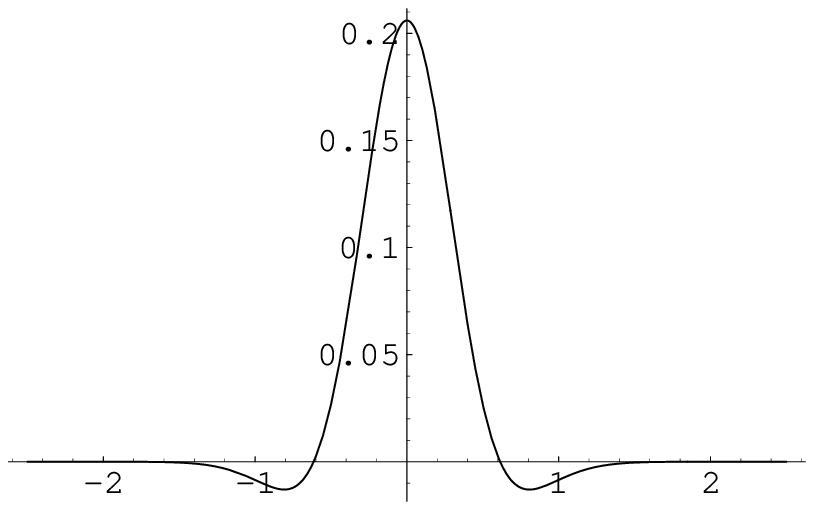}~2a~~
\includegraphics[width=4.9cm]{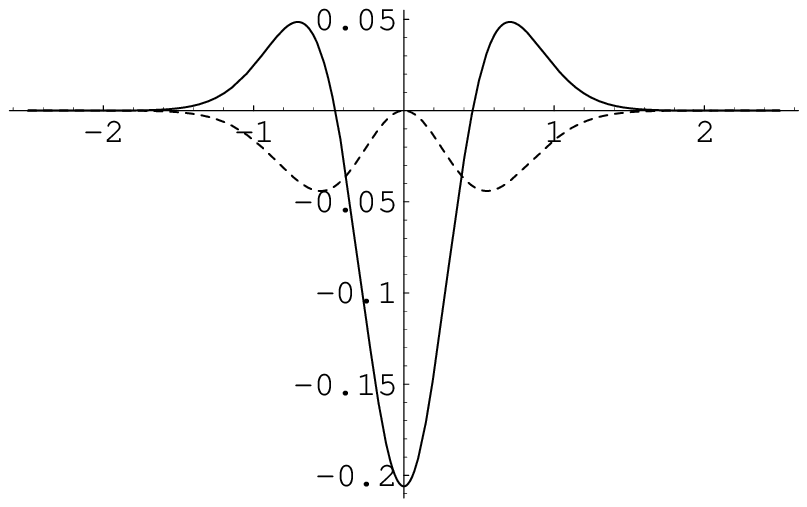}~2b~~
\includegraphics[width=4.9cm]{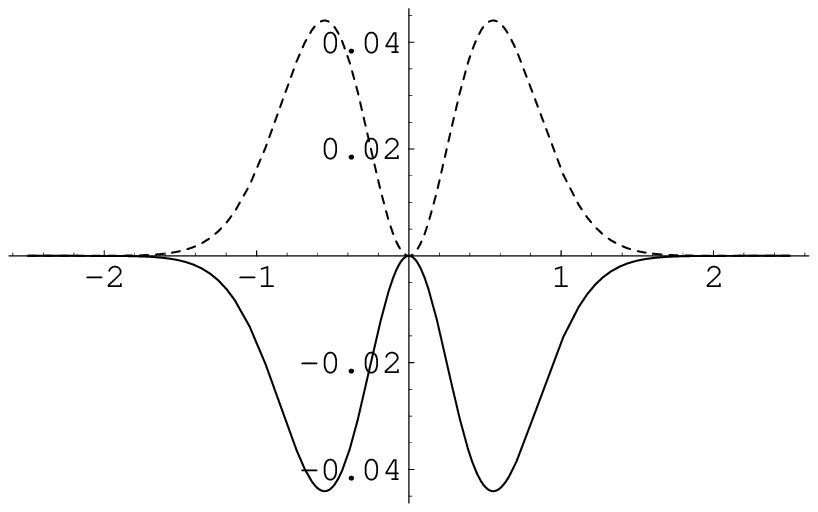}~2c~~
\caption{a) The potential energy $E_p$; b) The nonlocal terms
$E_{nl_1}$ (firm line) and   $E_{nl_2}$ (dashing line); c)
${\cal{E}}_p$ (dashing line) and ${\cal{E}}_k$ (firm line). 1a,1b
and 1c  correspond to $k=0.8,$ $\beta =1/4$, and 2a, 2b and 2c to
$k=0.2,$ $\beta =1/4$ \label{fig:potenexact.eps}
}
\end{figure}

On the solution (\ref{P-sol}) we can write explicitly $E_p$,
$E_{nl_1}$ and $E_{nl_2}$ as functions of  time. These functions
are presented on Fig.\ref{fig:potenexact.eps} for particular value
of parameters $\beta=1/4,$  $k=0.8$ and $k=0.2$. Time dependence
of ${\cal{E}}_p$ and ${\cal{E}}_k$ is also presented on Fig.
\ref{fig:potenexact.eps}

From Fig.\ref{fig:potenexact.eps}.c  we see that ${\cal E}_p\geq0$
and ${\cal E}_k\leq 0$ for $0\leq t<\infty $. Since the  energy is
conserved and equal to 0 and the  pressure is  $2{\cal E}_k\leq
0$.

\begin{figure}
\includegraphics[width=6cm]{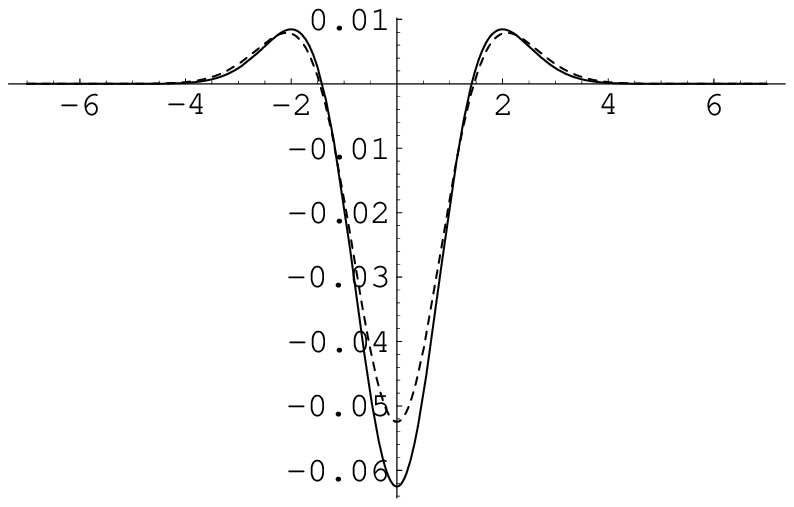}~1a~~~~~~
\includegraphics[width=6cm]{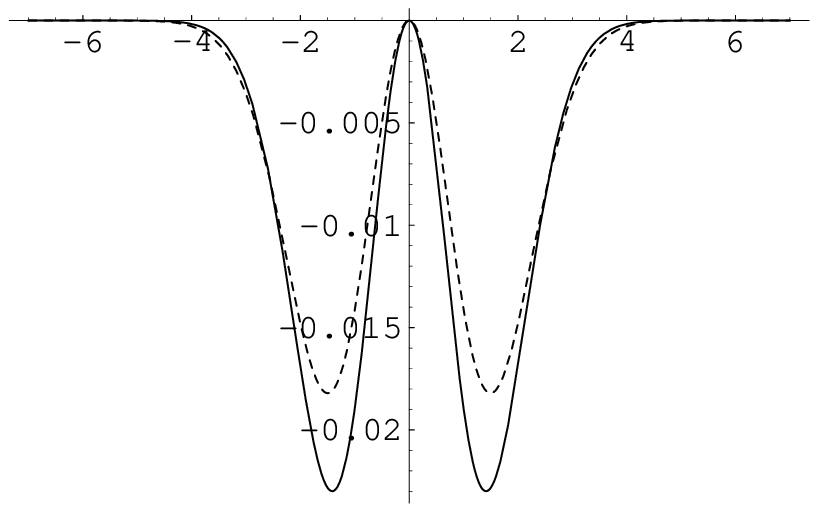}~1b~~\\
$~$\\
\includegraphics[width=6cm]{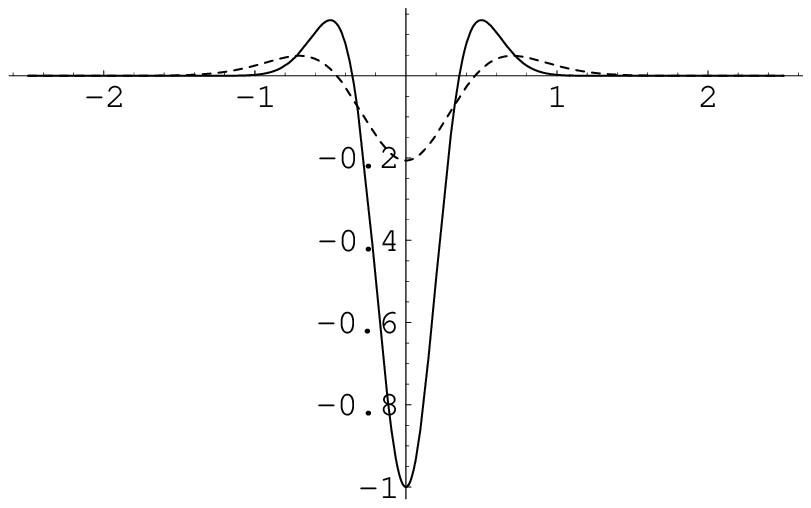}~2a~~~~~~
\includegraphics[width=6cm]{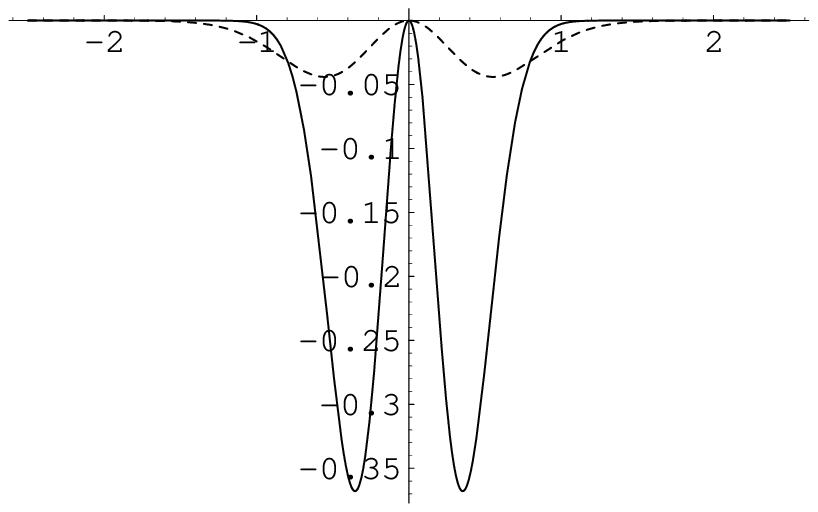}~2b~~
\caption{
On the left pictures (1a and 2a) the dashing lines
present the exact expressions $E_{nl_{1}}$ and the firm lines
present  the approximations to $E_{nl_{1}}$. On the right pictures
(1b and 2b) the dashing lines  present the exact expressions
$E_{nl_{2}}$ and the firm lines present  the approximations to
$E_{nl_{2}}$ . On 1a and 1b
 $k=0.8,$ $\beta =1/4$ and  on 2a and 2b   $k=0.2,$
$\beta =1/4$. \label{fig:approxforenl.eps}
}
\end{figure}

 Representing (\ref{En1}) and (\ref{En2}) as
\begin{equation}
\label{En1m} E_{nl_1}=\frac{\beta }{2}\int_{0}^{1}(e^{\beta
(1-\frac{\rho}{2} )\partial^2}\Phi)
 (e^{\frac{\beta }{2} \rho \partial^2} \partial^2 \Phi) d \rho
\end{equation}
and
\begin{equation}
\label{En2m} E_{nl_2}=-\frac{\beta }{2}\int_{0}^{1}((e^{\beta
(1-\frac{\rho}{2} )
\partial^2} \partial \Phi)) (e^{\frac{\beta }{2} \rho
\partial^2} \partial \Phi) d \rho.
\end{equation}
we can see that  in the approximation neglecting high order
derivatives one has
\begin{equation}
\label{Eka1} E_{nl_2}\sim -\frac{\beta }{2}\dot{\Phi} ^2,
\end{equation}
\begin{equation}
\label{Eka2} E_{nl_1}\sim \frac{\beta }{2}\Phi\ddot{\Phi} ,
\end{equation}
The discrepancy  in  these approximations are illustrated on the
fig.\ref{fig:approxforenl.eps}. We see that the discrepancy
becomes smaller when $k\to 1$.

In this approximation
\begin{equation}
\label{Ek} {\cal E}_k \cong -\frac{\beta }{2}\dot{\Phi} ^2
\end{equation}
\begin{eqnarray}
\label{Ep0} {\cal E}_p\cong
-\frac{1}{2}\phi^2+\frac{\sqrt{k}}{k+1}\Phi^{k+1}+ \frac{\beta
}{2}\Phi\ddot{\Phi}
\end{eqnarray}
Taking into account that in this approximation
\begin{equation}
\label{p2-P2} -\frac{1}{2}\phi^2\cong -\frac{1}{2}\Phi^2
-\frac{\beta }{2}\Phi\ddot{\Phi}
\end{equation}
we get
\begin{eqnarray}
\label{Ep} {\cal E}_p \cong
-\frac{1}{2}\Phi^2+\frac{\sqrt{k}}{k+1}\Phi^{k+1} \equiv V(\Phi)
\end{eqnarray}
%and using the previous notations,
%\begin{equation}
%\label{-effpot1} V=-{\cal V},
%\end{equation}
So we see that the approximated kinetic term ${\cal E}_k$
(\ref{Ek}) and potential ${\cal E}_p$ (\ref{Ep}) coincide up to
the sign with the kinetic and potential energy for the mechanical
problem (\ref{eq1-mech}).

\begin{figure}
\includegraphics[width=7cm]{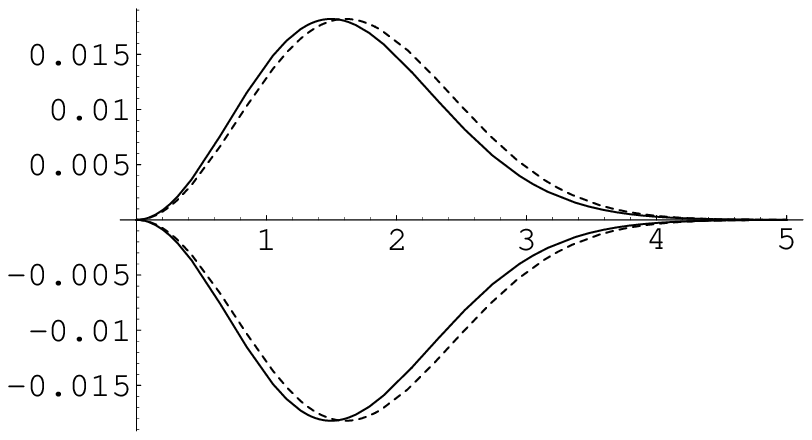}~a)~~
\includegraphics[width=7cm]{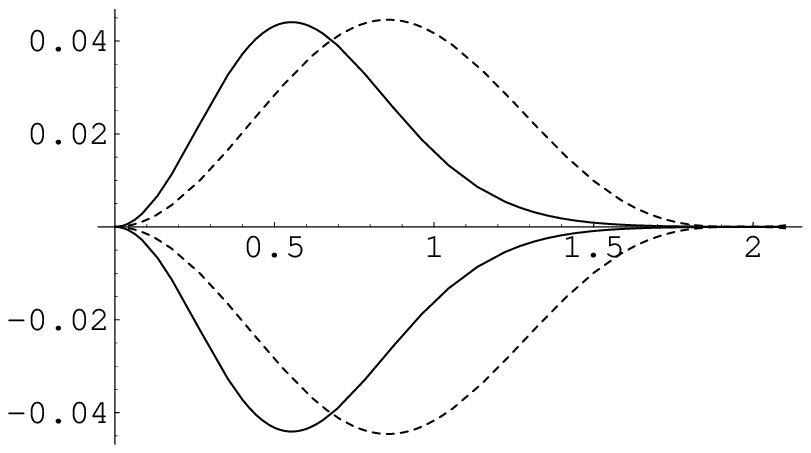}~b)~~
\caption{Local kinetic (dashing positive line) and potential
(dashing negative line) parts of the energy on zero-energy
solutions of eq.(\ref{eq1-mech}) as functions of time. Nonlocal
kinetic (firm negative line) and potential parts (firm positive
line) of the energy on the solutions (\ref{P-sol}); the picture
for  k=0.8 (a) and k=0.2 (b).
\label{fig:kin-pot-mech.eps}
}
\end{figure}

On Fig.\ref{fig:kin-pot-mech.eps} the kinetic and potential parts of
the energy for the local problem (\ref{eq1-mech}) as functions of
time  are compared
with ${ \cal E}_{k}$ and ${ \cal E}_{n}$
calculated on the solution (\ref{P-sol}). We see that for $k\to 1$
the kinetic and potential parts for local and nonlocal problems almost coincide
for time $0<t<T_k$.

%%%%%%%%%%%%%%%%%%%%%%%%%%%%%%%%%%%%%%%%%%%%%%%%%%%%%%%%%%%%%%%%%

\section{ Perturbations of Gaussian Lump} \label{sec:exp-pert}
\setcounter{equation}{0}

%%%%%\input lupm-exp-per-kinte34.tex
%%%%%%\newpage
\subsection{Perturbation by the friction}
Let us consider the integral equation with the constant friction
and time depending coupling constant
\begin{equation} \label{eqtm}
e^{\beta  \partial ^2_t+h\partial _t}\Phi_h=\sqrt{k}g(t)\Phi_h^k.
\end{equation}
For the case of a special time-depending coupling constant,
  $$g(t)=e^{-2h(1-k)t},$$
there is the following  lump solution
  \begin{equation}
\label{sol-If} \Phi_h(t)=e^{-\frac{b}{4\beta
}t^2+Bt},~~B=\frac{h}{2\beta }(-1-\sqrt{1+4\beta b})
\end{equation}
 Using the following integral representation
\begin{equation}
\label{int-h} e^{\beta  \partial ^2_t+h\partial_t}
\Phi(t)=\frac{1}{\sqrt{4\pi \beta }}\int
e^{-\frac{(t-t'+b)^2}{4\beta }}\Phi(t')dt'.
\end{equation}
one can check that (\ref{sol-If}) solves (\ref{eqtm})

Let us consider the  mechanical problem with a friction that
approximates  equation (\ref{exeq-h}),
\begin{equation}
\label{eq1-mech-h} \beta
\ddot{\varphi}(t)+h\dot{\varphi}(t)+\varphi(t)=g(t)\sqrt{k}\varphi(t)^{k},
\end{equation}
 The particle  with the zero energy
starts from  the  point
$\varphi_{max}=(\frac{2\sqrt{k}}{1+k})^{\frac{1}{1-k}}$ and moves
to $ \varphi_{0}=0$, but do not reach it.  The corresponding
trajectories
 are presented in the
fig.\ref{fig:potlump}.b for different $k$ and on
fig.\ref{fig:potlump}.c are presented the exact lumps
(\ref{P-sol}) for different $k$.

\begin{figure}
\includegraphics[width=3.5cm,angle=-90]{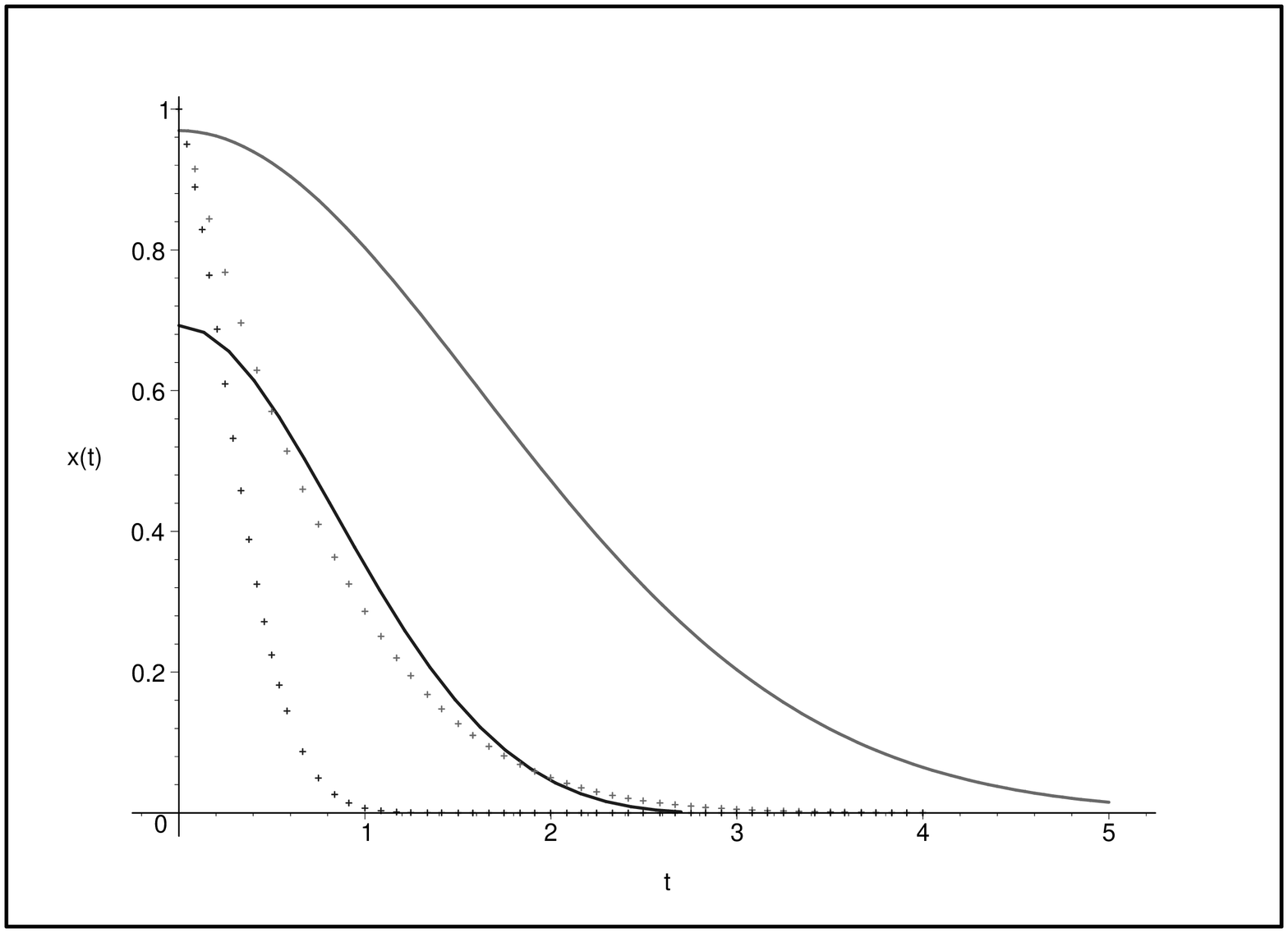}a)~~~~~
\includegraphics[width=3.5cm,angle=-90]{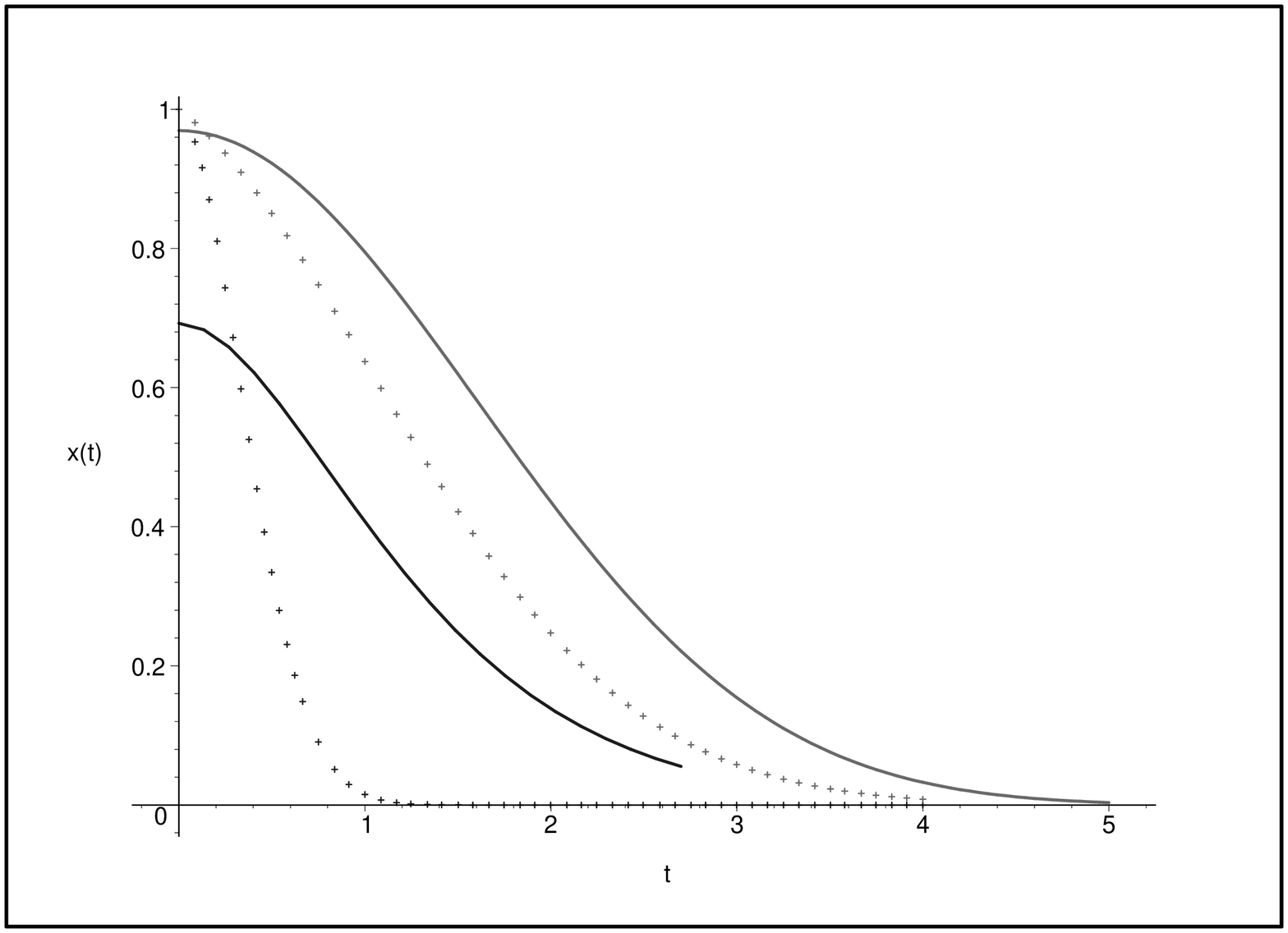}~b)
\caption{a) The numerical solution of equation (\ref{eq1-mech-h})
for k=0.2 (down solid line) and for $k=0.8$ (upper solid line) and
exact solution to integral equation (\ref{exeq-h}) for $\beta
=1/4$ and k= 0.2 (down dot  line) and k=0.8 (upper dot line),
h=0.5; b) the numerical solution of equation (\ref{eq1-mech-h})
for k=0.2 (upper solid line) and for $k=0.8$ (down solid line) and
exact solution to integral equation (\ref{exeq-h}) for $\beta
=1/4$ and k= 0.2 (down dot  line) and k=0.8 (upper dot line),
h=0.1
\label{fig:potlump-m}
}
\end{figure}

We see that the friction makes behavior of the solutions of the
nonlocal equation more closed to that of the
 corresponding mechanical problem.

\subsection{Perturbation by non flat metric}
Let us now consider an interaction of the theory
(\ref{action-exsol}) with the gravity
\begin{equation}
\label{action-E-M} S = \int \sqrt{-g}d^4x\,\left(
\frac{m_p^2}{2}R+\frac12 \phi^2-\frac{\sqrt{k}}{k+1}\Phi^{k+1}
\right)  \ .
\end{equation}
where
\begin{equation}
\label{met} \Phi=\exp(\frac \beta 2\Box_g) \phi,~~~~~~
\Box_g=\frac{1}{\sqrt{-g}}\pd _\mu \sqrt{-g}g^{\mu\nu} \pd _\nu
\end{equation}

On  space homogeneous configurations  in the Friedmann metric
(\ref{F-m})
 the field equation takes the form
 \begin{equation}
\label{eq-m} e^{\beta  {\cal D}}\Phi(t)=\sqrt{k}\Phi^k(t)
\end{equation}
where  $ {\cal D}=-\pd _t^2-3H(t)\pd _t $ and $H(t)=\dot{a}/a$,
$\dot{a}=\partial_t a$. The Einstein equations have the  form \bea
\label{n1}
3H^2&=&\frac{1}{m^2_p}~~\rho\\
\label{n2}
 H^2+2\ddot{a}/a&=&-\frac{1}{m^2_p}~~p\eea
with the energy and pressure densities are given by
\begin{equation}
\label{energy}
 \rho=-\frac{1}{2}(e^{-\frac{\beta }{2}}{\cal D})\Phi^2+
 \frac{\sqrt k}{k+1}\Phi^{k+1}+{\cal E}_{1}+{\cal E}_{2}
\end{equation}
\begin{equation}
\label{pressure} p=2{\cal E}_{2}
\end{equation}
where \beq \label{nonloc1} {\cal E}_{1}=-\frac{\beta
}{2}\int_{0}^{1} d\tau(e^{\frac{\beta }{2}\tau{\cal D}}\sqrt k
\Phi^k){\cal D}(e^{-\frac{\beta }{2}\tau{\cal D}}\Phi) \eeq \beq
\label{nonloc2} {\cal E}_2=-\frac{\beta }{2}\int_{0}^{1}
d\tau(e^{\frac{\beta }{2}\tau{\cal D}}\pd \sqrt k
\Phi^k)(e^{-\frac{\beta }{2}\tau{\cal D}}\pd \Phi) \eeq

Motivated by the flat case we make the following approximation
\beq \exp(\pd _t ^2+3H(t)\pd _t)\Phi \approx (~1+\pd _t
^2+3H(t)\pd _t~)~\Phi \eeq The corresponding Friedmann equations
have the form (\ref{energy}), (\ref{pressure}) with \bea
\label{1'}
\rho &=&-\frac12 \dot{\varphi}^2+V(\varphi),\\
p &=&-\frac12 \dot{\varphi}^2-V(\varphi) \eea and the equation for
$\varphi$  reads \beq \label{e.o.m.'}
\ddot{\varphi}+3H\dot{\varphi}=V'_{\varphi} \eeq Here $V$ is given
by (\ref{effpot1})

\begin{figure}
\includegraphics[width=4.7cm]{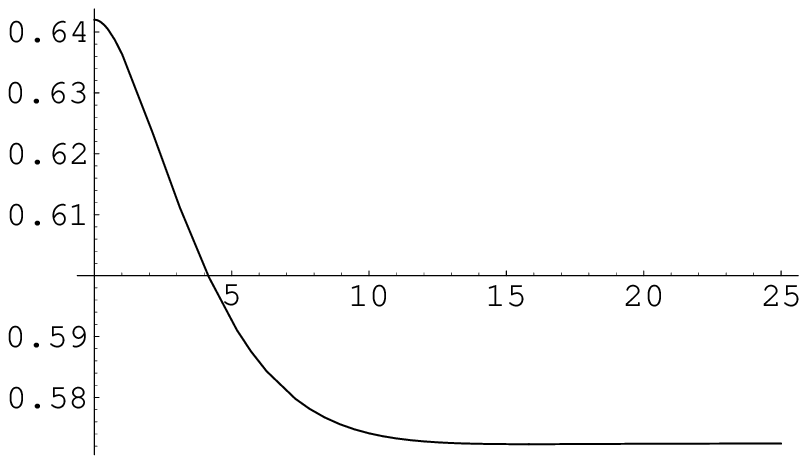}~a)~~
\includegraphics[width=4.7cm]{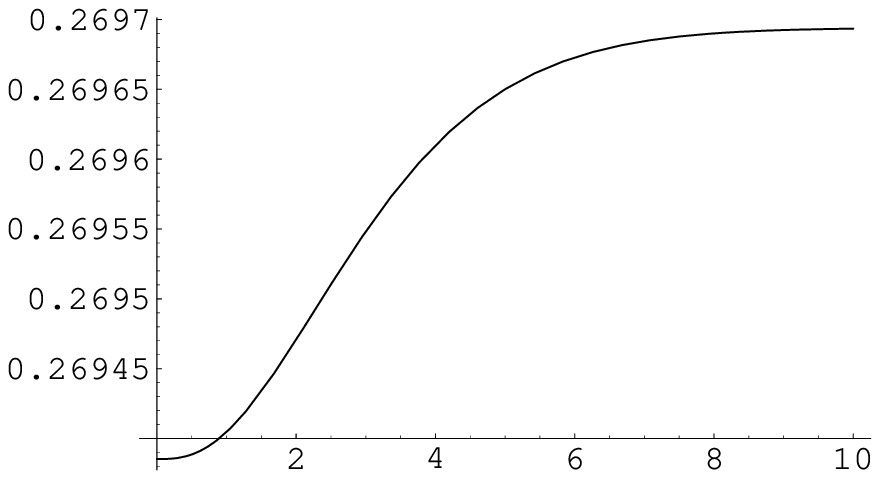}~b)~~
\includegraphics[width=4.7cm]{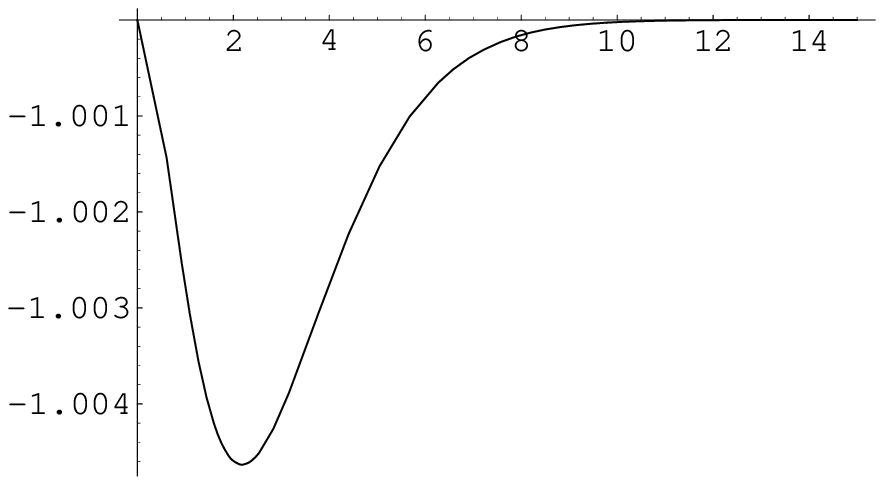}~c)~~
\caption{a) $\Phi$  as function on $t$ with the initial condition
$\varphi(0)=0.642$ and $\dot{\varphi}(0)=0$; b) function $H=H(t)$
as function on $t$; c) $w=w(t)$ as function on $t$.
\label{fig:H.eps}}
\end{figure}

An numerically solution of the system of equations (\ref{1'}) and
(\ref{e.o.m.'}) for the potential (\ref{effpot1}) is presented on
Fig.\ref{fig:H.eps}.

%We see that {\bf ADD MORE}

\subsection{Exact Gaussian Lump Solution in Friedman Metric}
In section \ref{sec:exp} we have analyzed the solution for the
phantom field with fractional interaction in the  Friedman metric.
Here we shall construct the potential which leads to exact
gaussian lump-type solution (\ref{P-sol}).
%We start from the action for a phantom scalar field in the
%gravitational background
%$$
%S=\int d^4 x \sqrt{-g}\left(M_{p}^2 R-\frac{1}{2}\phi \Box_g \phi
%-V(\phi) \right)
%$$
%where $g$ is the Friedman metric (\ref{}).
It is known that if one has an explicit  solution for the
space-homogeneous scalar field then one can reconstruct the form
of potential\cite{DW,AKV}. To realize this one  assumes that the
Hubble parameter $H(t)$ has the special form
\begin{equation}
\label{Hsup-egl} H(t)=W(\phi(t))
\end{equation}
where $W$ is a function called superpotential. Substituting the
anzats (\ref{Hsup-egl}) in the Friedmann equation
$\dot{H}=\dot{\phi}^2/2m_p^2$ we get
\begin{equation}
\label{supeqs1-egl}
 2m_{p}^{2} W'=\dot{\phi}
\end{equation}
where prime denotes derivatives in $\phi$. For the gaussian lump
solutions we have
\begin{equation}
\label{frid-egl} \dot{\phi}=-2 \phi\sqrt{-\xi ~ ln \phi},
\end{equation}
that gives the equation
\begin{equation}
\label{eq} 2M_{p}^{2} W'=-2 \phi\sqrt{-\xi ~ ln \phi},
\end{equation}
from which we reconstruct  the superpotential
\begin{equation}
\label{W-exp} W(\phi)=-\frac{1}{2M_p^2} \phi ^2  \sqrt{-\xi \log
\phi} +\frac{\sqrt{2\pi \xi }}{8m^2_p}
       \mbox{erf}(\sqrt{-2\log \phi })
       \end{equation}
The potential is obtained from the second Friedmann equation,
$V(\phi)=1/2\dot{\phi}^2 +3m^2_pW^2$, and
\begin{equation}
\label{V-exp} V(\phi)=-2\xi \phi ^2\,\log \phi  +
  \frac{3\,\xi}{64\,m^2_p} \left( {\sqrt{2\,\pi }}\,
          {\mbox erf}
            (\sqrt{-2\log \phi }) -
         4\,{\phi }^2\,{\sqrt{-\log (\phi )}} \right)^2
\end{equation}

\begin{figure}
\includegraphics[width=4.3cm]{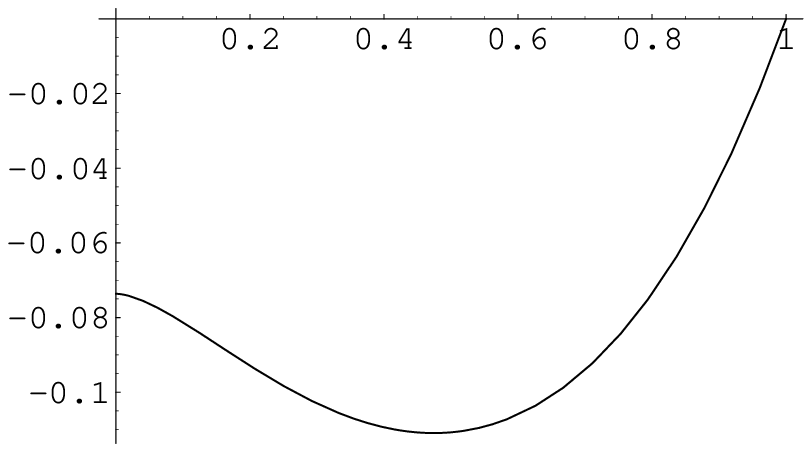}~a)
\includegraphics[width=4.3cm]{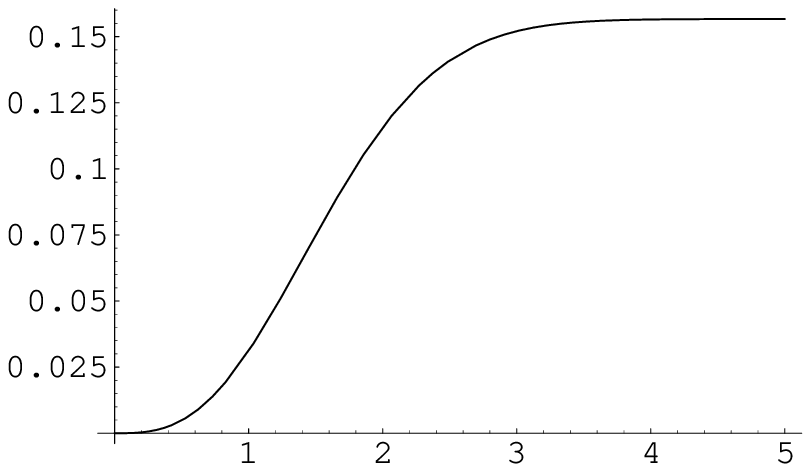}~b)
\includegraphics[width=4.3cm]{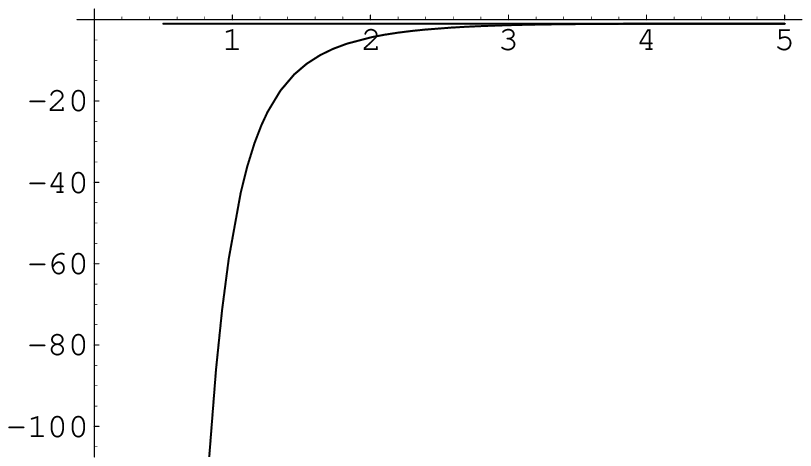}~c)
\caption{a) The overturn potential $-V$,  b) $H=H(t)$ as function
on $t$; c) $w=w(t)$ as function on $t$  and the line $w=-1$
\label{fig:frid-egl.eps}}
\end{figure}

The Hubble parameter has the form
\begin{equation}
\label{H-exp} H(t)=-\frac{1}{8\,m^2_p}\left(4\,t\,\xi e^{-2\xi t^2
} -\sqrt{2\,\pi \xi}\,
      erf(\sqrt{2\xi }\,t)\right)
\end{equation}
The asymptotic behavior of the Hubble parameter and the scale
factor are
\begin{equation}
\label{H-a-exp} H(t)\sim H_0=\frac{\sqrt{2 \pi \xi}}{8
m_p^2},~~~~~ a(t)\sim e^{ H_0t}~~~for~large~t
\end{equation}
The state parameter
 has the following explicit form
\begin{equation}
\label{w-ext} w(t)=-1 - \frac{256\xi t^2 e^{2\xi t^2 }m^2_p}
 {3\,{\left( -4\,t\,{\sqrt{\xi }} +
         e^{2\,t^2\,\xi }\,{\sqrt{2\,\pi }}\,
          erf({\sqrt{2}}\,t\,
            {\sqrt{\xi }}) \right) }^2}
\end{equation}
The shapes of the potential, the Hubble parameter and the state
parameter are presented at (Fig.\ref{fig:frid-egl.eps}). It is
interesting to note that the next to leading term in the
asymptotic of (\ref{w-ext}) at large $t$
\begin{equation}
\label{w-as-ext} w(t)\sim -1 - \frac{128\,m^2_p\,t^2\,\xi
\,e^{-2\xi t^2}}{3\pi }
\end{equation}
is  proportional to $m^2_p$ meanwhile from the general formula
(\ref{omega-H}) one can expect that the deviation from the
limiting value -1 could be proportional to $1/m^2_p$. This is
caused by a presence of a dependence of $m^2_p$ in $H_0\sim
1/m^2_p$ and independence of $\dot{\phi}$ on $m^2_p$.

%%%%%%%%%%%%%%%%%%%%%%%%%%%%%%%%%%%%%%%%%%%%%%%%%%%%%%%%%%%%%%%%%
\section{Hyperbolic Lump}
\label{sec:sech}
\setcounter{equation}{0}
\subsection{Numerical construction of the potential}
Here we will be interested in the scalar SFT-type nonlocal models
 described by the following action
\begin{equation}
\label{Ssech} S=\int d^d x \left[\frac{1}{2}\phi^2-
U(\Phi)\right],
\end{equation}
where $U$ is a polynomial  and $\Phi$ the smoothed field given by
(\ref{Pp}). Action (\ref{Ssech}) leads to the equation of motion
which on the space homogeneous configurations  reads
\begin{equation}
\label{eomsech}
 e^{\beta  \partial _t^2}\Phi=U(\Phi)
\end{equation}
We assume that $U$ has a form
\begin{equation}
\label{W}
U(\Phi)=U_{\vec{\alpha}}(\Phi)\equiv\sum_{n=1}^{N}\frac{\alpha_n}{n+1}
\Phi^{n+1},
\end{equation}
%\begin{figure}
%\centering
%\includegraphics[width=7cm]{sech2.eps}
%\caption{Функция $\varphi_{0}(t)=\sech^2(t)$.} \label{fig:sech2}
%\end{figure}
We choose a function
\begin{equation}
\label{sec} \Phi_{0}(t)=\sech^2(t)
\end{equation}
and find the set  of coefficients $\{\alpha_n\}$ from a
requirement that that  a discrepancy
\begin{equation}
\label{deltaN} \delta_N(\vec{\alpha})=e^{\beta  \partial
_t^2}\Phi-U'(\Phi)
\end{equation}
is small in $L_2$-norm for fixed $N$, where $N$ is a number of
terms in the series (\ref{W}). For numerical calculations we use a
representation of the left hand side of (\ref{eomsech}) in term of
the integral operator
\begin{equation}
\label{int} e^{\beta \partial _t^2}\Phi (t)\equiv
K[\Phi](t)=\frac{1}{\sqrt{4\pi
\beta}}\int\limits_{-\infty}^{\infty}e^{-\frac{(t-\tau)^2}{4\beta
}} \Phi(\tau)d \tau,
\end{equation}
and take $\beta=\frac{1}{4}$. We denote the R.H.S. of
(\ref{eomsech}) as a  force  $F$
\begin{equation}
\label{FN-op} F_{\vec{\alpha}}[\Phi]\equiv F_{\alpha_N}[\Phi]=\sum_{n=1}^{N}\alpha_n
\Phi^n,
\end{equation}

\begin{figure}
\includegraphics[width=4.3cm]{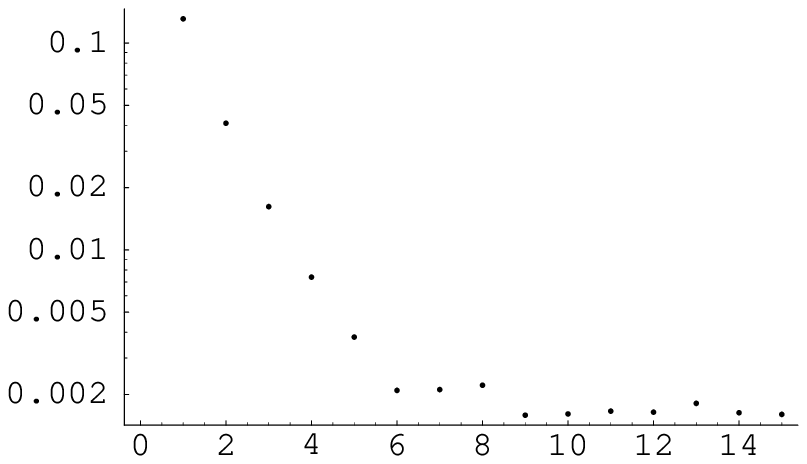}~a)
\includegraphics[width=4.3cm]{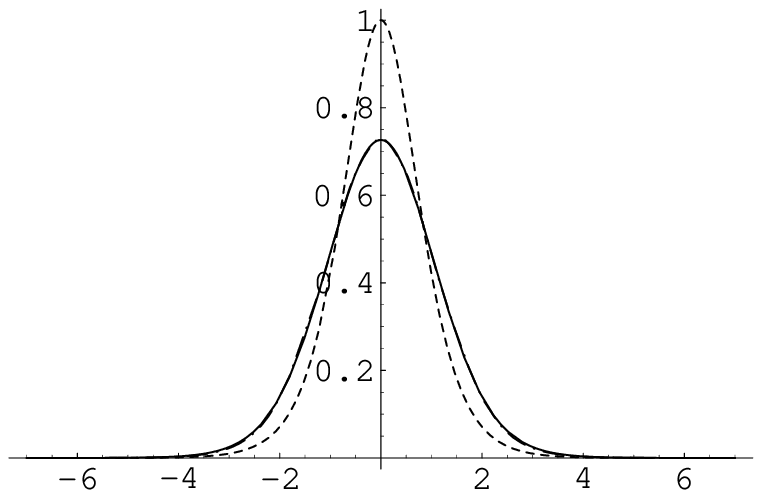}~b)
\includegraphics[width=4.3cm]{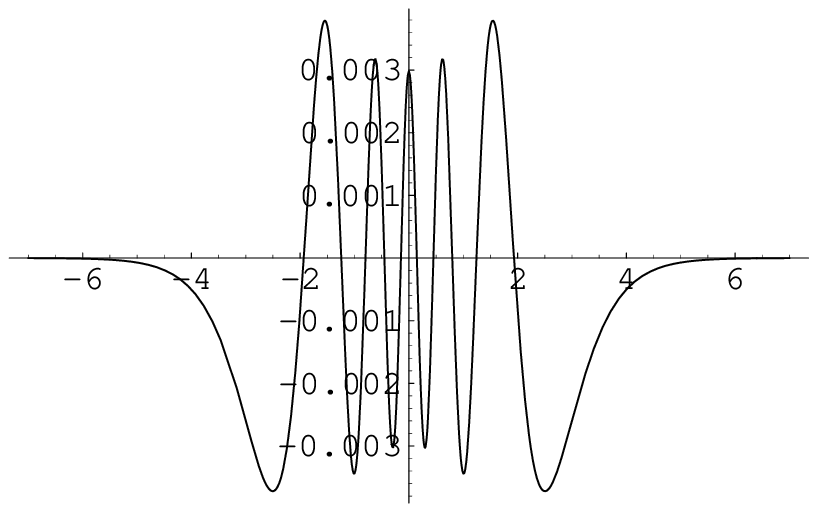}~c)
\caption{a) Minimum of the absolute value of the function of
discrepancy  $\delta_N(\vec{\alpha})$ (\ref{deltaN}) (in
\textit{the logarithmic} scale at pintle) depending on  number of
terms  $N$ is presented; b)$\Phi_{0}(t)=\sech^2(t)$ is presented
by dash line, function  $K[\Phi_0](t)$ is presented by the firm
line and function $F_{\alpha_5}[\varphi_0]$ is presented by
dotdashing line  $N=5$, coefficients $\alpha_5$ are collected on
the Table; c) function
$\delta_5(\alpha_5)=K[\varphi_0]-F_{\alpha_5}[\varphi_0]$.
\label{fig:kff5}}
\end{figure}

Note that numerical calculations show that we cannot make the
discrepancy equal to zero for final $N$, this illustrate the
fig.{\ref{fig:kff5}}a).

Note that already for  $N=5$ the discrepancy  is rather small,
this illustrate the fig.{\ref{fig:kff5}}b). The coefficients
$\alpha$ for different $N$ are presented in the  Table
$\vec{\alpha}$
\begin{center}
{\small \begin{tabular}{|c|c|c|c|c|c|c|c|}
\hline\\
$N$ & $\alpha_1$ & $\alpha_2$ &$\alpha_3$ & $\alpha_4$ &
$\alpha_5$ & $\alpha_6$ &
$\alpha_7$\\
\hline 3& 1.896 &-2.2206 &1.0649 & -&-& -&-
 \\
 \hline
5& 2.319 &-5.7831 &10.1891 & -9.1559&3.1602& -&-
 \\
\hline
6& 2.4309& -7.5998& 19.2683&-28.4428 & 21.4779& -6.4098&-  \\
\hline
12 & 2.4628 & -8.2965 & 23.6528 & -38.931 & 28.6269 & 0.287057 & -6.3996 \\
\hline
13 & 2.4643 & -8.3152 & 23.7106 & -38.965 & 28.5816 & 0.278636 & -6.3756 \\
\hline
14 & 2.4662 & -8.3356 & 23.7655 & -38.985 & 28.5391 & 0.256198 & -6.3702 \\
\hline
15 & 2.4602 & -8.2969 & 23.7264 & -39.024 & 28.5326 & 0.282346 & -6.3252 \\
\hline
\end{tabular}}
\end{center}
%\alpha_{11} & \alpha_{12}

It is interesting to note that as follows from the
fig.~\ref{fig:kff5}, there is a tendency of decreasing of the
minimum value of the discrepancy when the number of terms in the
series (\ref{W}) is increase, moreover the values of the low-order
coefficients are stabilized when $N$ is increase.

\subsection{Numerical solution of the integral equation}

We are going to solve the equation
\begin{equation}
\label{exeq} K[y](t)=F_{\alpha_N}[y](t),
\end{equation}
where the right hand side  is given by  $F_{\alpha_N}$, that
minimized the discrepancy for the function (\ref{sec}) A solution
of this equation we find using the following iteration procedure
\begin{equation}
\label{iterpr} y_{n+1}=y_{n}+\lambda
(K[y]-F_{\alpha_N}[y]),~~\lambda=0.01,
\end{equation}
and as the zero approximation we take (\ref{sec}), i.e. $
y_{0}=\sech^2(t) $

\begin{figure}
\includegraphics[width=4.3cm]{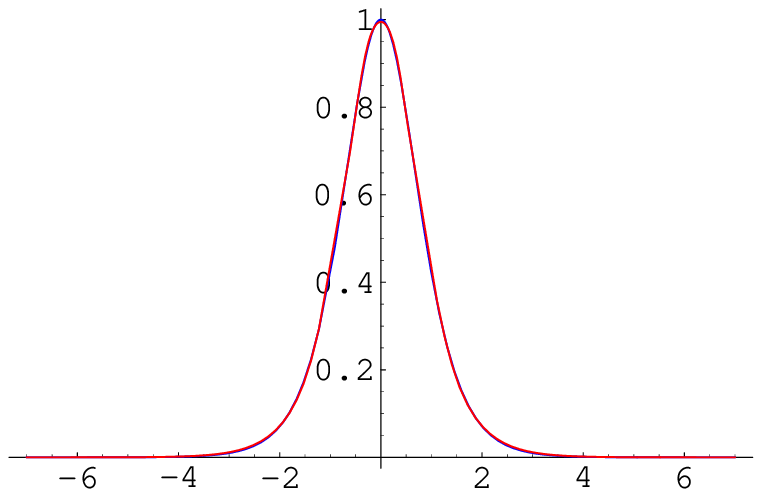}~a)
\includegraphics[width=4.3cm]{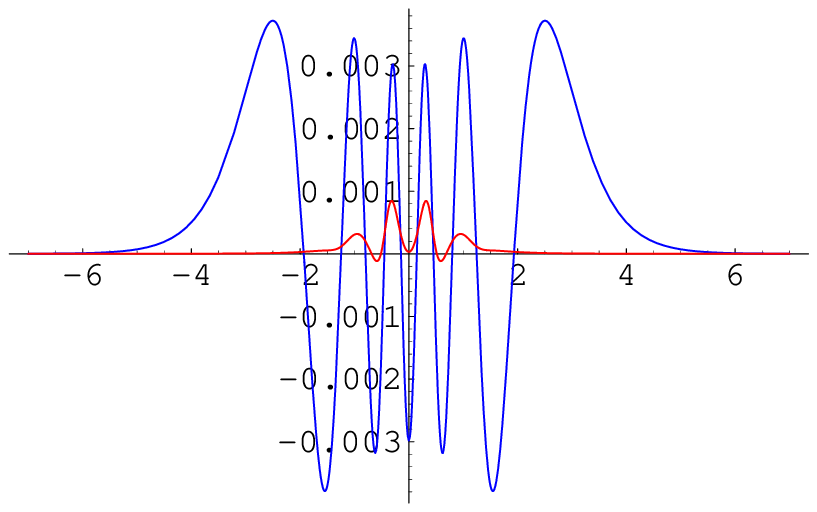}~b)
\includegraphics[width=4.3cm]{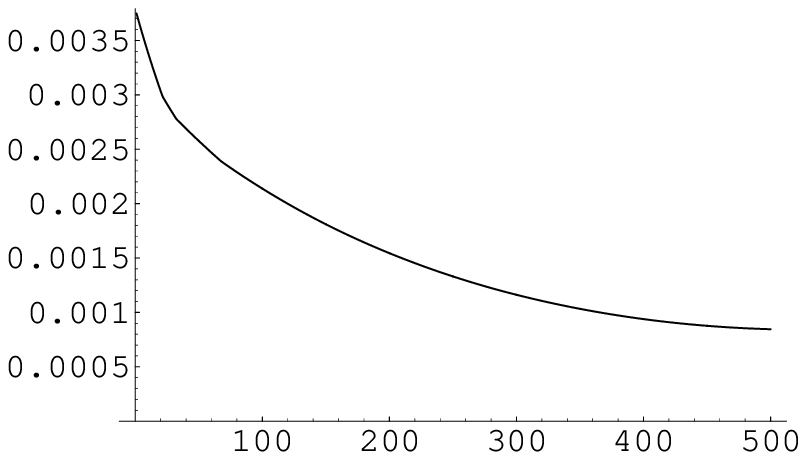}~c)
\caption{a) Zero iteration $y_{0}=\sech^2(t)$ is depicters by firm
line
 and five hundredth 500 by dashing line($N=5$);
 b)The discrepancy for initial iteration (the line with amplitude $0.003$)
 and for solution $y(t)$ of equation (\ref{exeq})(the line with amplitude $0.001$);
c)The dependence for descrepansy (\ref{deltanl}) as funcion on
number of iteration when the number of terms in the series $N=5$.
} \label{ris.4}
\end{figure}

Numerical calculations show that this iteration procedure
converges. A character of a convergent of the iteration procedure
is natural to specify by the a maximum of the discrepancy
\begin{equation}
\label{deltanl}
\Delta[y_n]=\max_{t}{\left|K[y_n](t)-F_{\alpha_N}[y_n](t)\right|}
\end{equation}
As we see that $\Delta[y_n]$ becomes smaller for a large number of
iteration $n$ (when the number of terms in the series $N=5$),
however as we can see from Fig.~\ref{ris.4}~c) at $n\sim 500$
$\Delta[y_n]$ goes to a ``plateau"  and after decreases slowly.
The same situation takes place when we consider the case when the
number of terms in the series $N=7,14,21,..etc$.

\subsection{Mechanical problem}
The mechanical problem that corresponds to (\ref{exeq})
 has the form
\begin{equation}
\label{solmech-exp} \beta  \ddot{
\varphi(t)}=F_{\alpha}[\varphi]-\varphi.
\end{equation}
The potential  for this problem is $V_{ot}(\varphi)$
\begin{equation}
\label{sech-pot}
 V_{ot}(\varphi)=\frac{1}{2}\varphi^2-U_{\vec{\alpha}}(\varphi)
\end{equation}
 $V_{ot}(\varphi)$ is
the overturn version of the potential $V$ given by (\ref{pot})
with $U$ given by (\ref{W}).

\begin{figure}
\includegraphics[width=3.5cm]{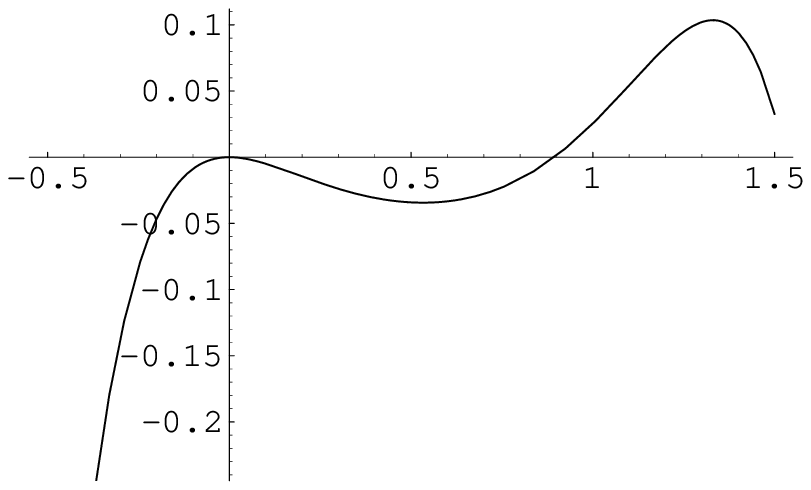}a)~~~
\includegraphics[width=3.5cm]{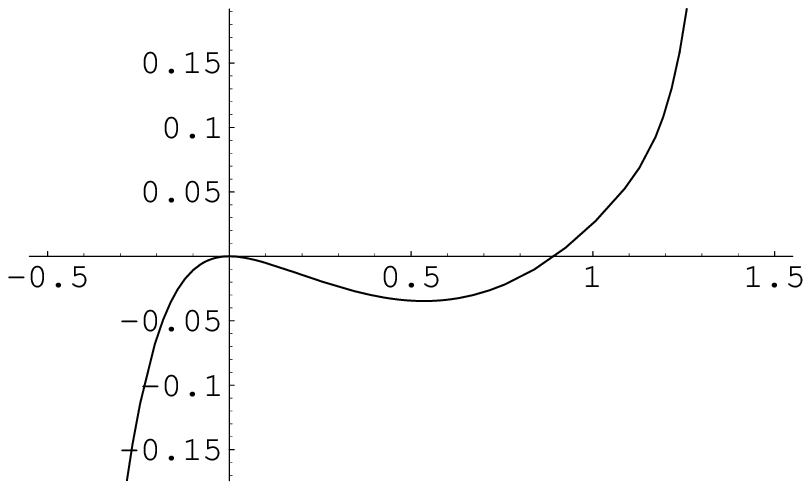}b)~~~
\includegraphics[width=3.5cm]{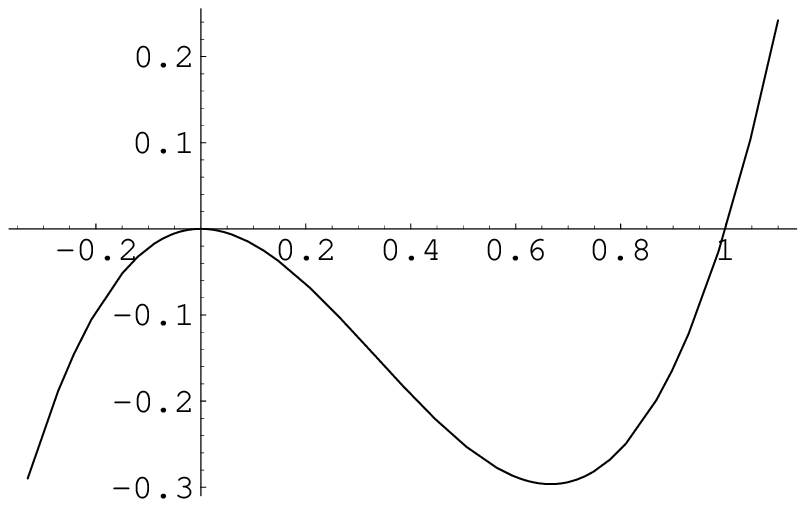}c)~
\caption{The overturn potential (\ref{sech-pot})  when N is equal:
a) $N=5$ and b) $N=14$; c) the overturn potential $-V_0$}
\label{fig:pot5}
\end{figure}

The overturn potentials for N=5 and N=14 are presented on
fig.\ref{fig:pot5}. It is interesting to note that  global forms
of corresponding potentials are different for different $N$ see
fig.\ref{fig:pot5}, but in all considered cases there  are local
maximum of the potential in the zero field configuration  and
points of local minimum.

Starting from the zero-energy bounce point,
 $\varphi(0)=\varphi_0$, where $V(\varphi_0)=0$,
with zero velocity, $\dot{\varphi}(0)=0$,
 the particle moves to  the local maximum
of the potential at the  point $\varphi=0$ during  infinite time.
For the initial data $\varphi(0)=\varphi_1<\varphi_0$ and
$\dot{\varphi}(0)=0$ the particle performs a periodic motion. For
$\varphi(0)=\varphi_0$ and $\dot{\varphi}(0)\neq 0$ the motion is
unbounded.

On Fig.\ref{fig:pot5}.c for a comparison we present  the potential
$-V_0$. As it is known (see also  the  subsection (4.5)) the lump
(\ref {sec}) is a zero-energy solution of the mechanical problem
with the phantom kinetic term and potential $V_0$ given by
(\ref{pot-sech}).

\subsection{Energy  and pressure}
In this subsection similar to the subsection 2.3 we consider the
energy and the pressure of the nonlocal problem (\ref{exeq}) and
compare them with energy and pressure of the corresponding
mechanical problem (\ref{solmech-exp}). Equation (\ref{exeq}) has
the conserved energy functional
\begin{equation}
\label{E-m} E=E_{p}+E_{nl_1}+E_{nl_2},
\end{equation}
where
\begin{equation}
\label{Ep1-m} E_{p}=-\frac{1}{2}\phi^2+U(\Phi),
\end{equation}
\begin{equation}
\label{En1-m} E_{nl_1}=\frac{\beta
}{2}\int_{0}^{1}(e^{-\frac{\beta}{2} \rho
\partial^2} U'(\Phi))
 (e^{\frac{\beta }{2} \rho \partial^2} \partial^2 \Phi) d \rho
\end{equation}
and
\begin{equation}
\label{En2-m}
E_{nl_2}=-\frac{a}{2}\int_{0}^{1}((e^{-\frac{\beta}{2} \rho
\partial^2} \partial U'(\Phi) (e^{\frac{\beta}{2} \rho
\partial^2} \partial \Phi) d \rho.
\end{equation}

The pressure
\begin{equation}
\label{P} P=E_{nl_2}-E_p-E_{nl_1}
\end{equation}
has the representation  (\ref{P}) and
\begin{equation}
\label{pressure-m} P=-a\int_{0}^{1}(e^{(2-\rho)\frac{\beta}{2}
\partial^2}\partial \Phi) (e^{\frac{\beta}{2} \rho \partial^2}
\partial \Phi) d \rho
\end{equation}

\begin{figure}
\includegraphics[width=5cm]{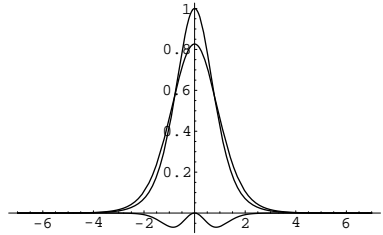}
\caption{a)The lump (\ref{sec}) and the corresponding smothered
function (lowest positive line in the picture) and the pressure
(\ref{pressure-m}) for $N=14$ (negative line).}
\label{fig:pressure-exp}
\end{figure}

%%%%%%%%%%%%%%%%%%%%%%%%%%%%%%%%%%%%%%%%%%%%%%%%%%%%%%%%%%%%%%%%%
%\newpage

\subsection{Exact Lump Solution in Friedmann Metric}
\begin{figure}
\includegraphics[width=4.3cm]{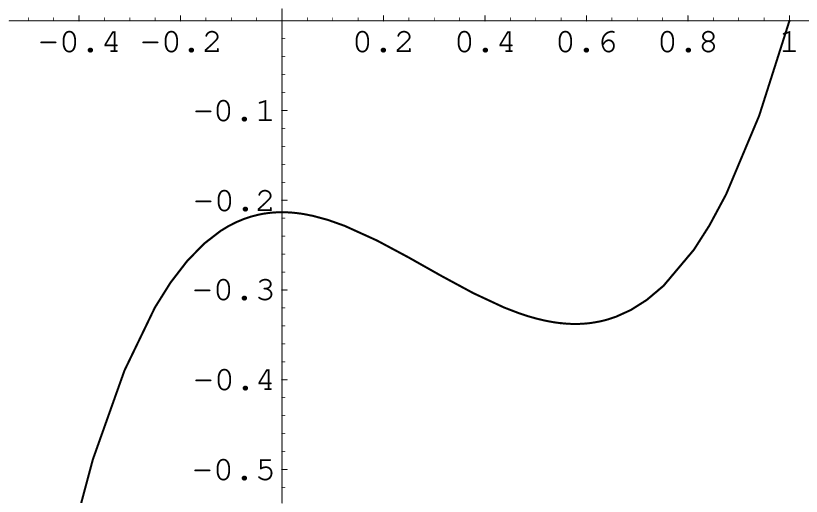}a)
\includegraphics[width=4.3cm]{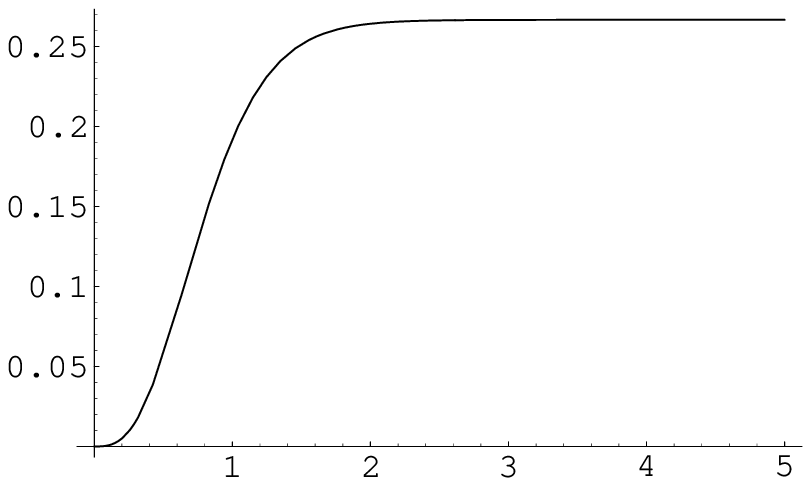}b)
\includegraphics[width=4.3cm]{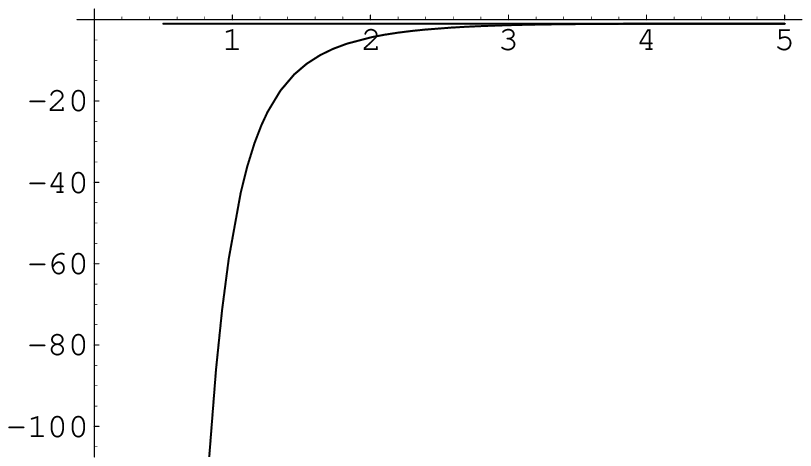}c)
\caption{a)The overturn potential $-V$,  b) $H=H(t)$ as function
on $t$; c) $w=w(t)$ as function on $t$} \label{fig:Vfrid.eps}
\end{figure}

In the full analogy with what has been done in sect.3.3 we can
easily find a potential for which the function (\ref{sec}) is a
solution to the Friedmann equations. It has the form
\begin{equation}
\label{pot-sec} V(\phi)=2\,\left( 1 - \phi  \right) \,{\phi }^2 -
  \frac{4\,{\left( -1 + \phi  \right) }^3\,
     {\left( 2 + 3\,\phi  \right) }^2}{75\,{{m_p}}^2}
\end{equation}
The shape of the potential is presented at
fig.\ref{fig:Vfrid.eps}.a.

The Hubble parameter has the form
\begin{equation}
H(t)=\frac{2\,\left( 4 + \cosh (2\,t) \right) \,
    {\sech(t)}^2\,{\tanh (t)}^3}{15\,
    {{m_p}}^2}
\end{equation}
and its shape is presented on (Fig.\ref{fig:Vfrid.eps}.b). $H(t)$
and the scale factor $a(t)$ have the following  asymptotic
behavior at large $t$
\begin{equation}
\label{H-as-s} H(t)\sim \frac{2}{15 m_p^2},~~~~~a(t)\sim
e^{\frac{4}{15 m_p^2}t}~~~ for~~t\to \infty
\end{equation}

 The state parameter is given by
 the following explicit formula

\begin{equation}
w(t)=-1-\frac{75\,m_p^2\,{\coth (t)}^4}
  {{\left( 4 + \cosh (2\,t) \right) }^2}
\end{equation}
and we see that for large times $w(t)$ goes to $-1$.

%%%%%%%%%%%%%%%%%%%%%%%%%%%%%%%%%%%%%%%%%%%%%%%%%%%%%%%%%%%%%%%%%
%\newpage
%\section{Conclusion Remarks}
%\label{sec:conc} \setcounter{equation}{0}
$$~$$

{\large {\bf Acknowledgements}}

We would like to thank Ya. Volovich for help with numerical
calculations. This work is supported in part by RFBR grant
05-01-00578 and INTAS grant 03-51-6346. I.A. is also supported by
the grant for scientific schools.  L.J. is also financially
supported by the fellowship of the ``Dynasty" Foundation and
International Center for Fundamental Physics in Moscow, by the
grant of the ``Russian Science Support Foundation",
and by the personal grant from Swedish Institute.
$$~$$

%%%%%%%%%%%%%%%%%%%%%%%%%%%%%%%%%%%%%%%%%%%%%%%%%%%%%%%%%%%%%%%%%
%\newpage


\begin{thebibliography}{99}


\bibitem{Sen} A.~Sen, JHEP {\bf 0204}, 048 (2002), {\it Rolling Tachyon},
hep-th/0203211;
 A.~Sen, JHEP {\bf 0207}, 065 (2002), {\it Tachyon Matter},
hep-th/0203265; {\it Tachyon Dynamics in Open String Theory},
hep-th/0410103.

\bibitem{GG} G.W. Gibbons, {\it Thoughts on Tachyon Cosmology},
Class.Quant.Grav. 20 (2003) S321-S346, hep-th/0301117.

\bibitem{FS} Andrei V. Frolov, Lev Kofman, Alexei A. Starobinsky,
    {\it Prospects and problems of tachyon matter cosmology},
    Phys.Lett.B545 (2002) 8-16,
    hep-th/0204187;

    Gary N. Felder, Lev Kofman, Alexei Starobinsky,
    {\it Caustics in tachyon matter and other Born-Infeld scalars}.
    JHEP 0209 (2002) 026,  hep-th/0208019.
%%%%%%%%%%%%%%
%%%%%%%%%%%%%%%    Branes scenario of accelaration
%%%%%%%%%%%%%%%

\bibitem{brane} V. Sahni, Y. Shtanov, {\it Brane World Models of Dark
    Energy}
    JCAP 0311:014,2003, astro-ph/0202346;

Ph. Brax, C. van de Bruck and A.-C. Davis, {\it Brane world
cosmology}, Rept.Prog.Phys.67:2183-2232, hep-th/0404011;


B. McInnes,  {\it  The phantom divide in string gas cosmology},
hep-th/0502209.

\bibitem{IA1}
 I.Ya.Aref'eva,
 {\it Nonlocal String Tachyon as a Model for Cosmological Dark Energy},
 astro-ph/0410443.

\bibitem{Perlm} S.J. Perlmutter et al., {\it Measurements of Omega and
Lambda from 42 High-Redshift Supernovae}, \APJ {\bf 517} (1999)
565, astro-ph/9812133.
\bibitem{Riess}  A. Riess et al.,
{\it Observational Evidence from Supernovae for an Accelerating
Universe
 and a Cosmological Constant},   \AJ {\bf 116} (1998) 1009, astro-ph/9805201;\\
A. Riess et al.,
{\it Supernova Search Team},   \AJ {\bf 607} (2004) 665, astro-ph/0402512.

\bibitem{knop} R.A.Knop et al., {\it New constraints on $\omega_m$, $\omega_\lambda$, and
w from an independent set of eleven high - redshift supernovae
observed with HST}, astro-ph/0309368.

\bibitem{Tegmark} M. Tegmark al., {\it The 3-d power spectrum of galaxies
    from the SDSS},  \APJ 606 (2004) 702-740,  astro-ph/0310723.

\bibitem{Spergel}
 D. N. Spergel et al.,  {\it First Year Wilkinson Microwave Anisotropy Probe (WMAP) Observations: Determination of
 Cosmological Parameters}, \APJ Suppl. {\bf 148} (2003) 175, astro-ph/0302209.
%%%%%%%%%%%%%%%%%%%%%
%%%%%%%%%%%%%%% -1<w<0, review

\bibitem{Sachni} V. Sahni,  {\it Dark Matter and Dark Energy},
astro-ph/0403324.

\bibitem{Frampton} P. Frampton, {\it Dark energy - a pedagogic review},
hep-th/0409166.
%%%%%%%%%%%%%%%%%%%%%
%%%%%%%%%%%%%%% w<-1
\bibitem{Caldwell}
Caldwell R R,  {\it A Phantom Menace? Cosmological consequences of
a dark energy component with super-negative equation of state},
\PL B {\bf 545} (2002) 23, astro-ph/9908168.

\bibitem{Innes}
McInnes B,  {\it The dS/CFT Correspondence and the Big Smash},
JHEP {\bf 0208} (2002) 029, hep-th/0112066.

\bibitem{Caldwell03}
Caldwell R R, Kamionkowski M and Weinberg N N,  {\it Phantom
Energy and Cosmic Doomsday}, \PRL {\bf 91} (2003) 071301,
astro-ph/0302506.

\bibitem{woodard}
V.K. Onemli, R.P. Woodard, {\it Super-Acceleration from Massless,
Minimally Coupled $\phi^4$}, Class.Quant.Grav. 19 (2002) 4607,
gr-qc/0204065; V. K. Onemli, R. P. Woodard {\it Quantum effects
can render w<-1 on cosmological scales}, Phys.Rev. D70 (2004)
107301, gr-qc/0406098.

\bibitem{carroll03}
S.M.Carroll, M.Hoffman  and M.Trodden, {\it Can the dark energy
equation-of-state parameter w be less than -1? },  \PR D {\bf 68}
(2003) 023509, astro-ph/0301273.

\bibitem{Melchiorri}
A.Melchiorri, L.Mersini, C.J. Odman  and Trodden M, {\it The State
of the Dark Energy Equation of State}, \PR D {\bf 68} (2003)
043509, astro-ph/0211522.





\bibitem{Padmanabhan-rev}
T.Padmanabhan, {\it Cosmological Constant - the Weight of the
Vacuum}, Phys.Rept. 380 (2003) 235-320, hep-th/0212290.


\bibitem{Bo}
Bo Feng, Xiulian Wang, Xinmin Zhang, {\it Dark Energy Constraints
from the Cosmic Age and Supernova}, astro-ph/0404224; Bo Feng,
Mingzhe Li, Yun-Song Piao, Xinmin Zhang, {\it Oscillating Quintom
and the Recurrent Universe}, astro-ph/0407432.


\bibitem{Solod}
S. Nojiri and S.Odintsov, {\it Properties of singularities in
(phantom) dark energy universe}, Phys.Rev.D71, 063004, 2005,
hep-th/0501025.

\bibitem{kitaj}
W.Fang, H.Q.Lu, Z.G. Huang and K.F..Zhang, {\it Phantom Cosmology
with Born-Infeld Type Scalar Field},  hep-th/0409080.


 \bibitem{AKV}
 I.Ya.Aref'eva, A.Koshelev and S.Vernov,
 {\it Exactly Solvable SFT Inspired Phantom Model},
 astro-ph/0412619.


%%%%%%%%%%%%%%%%%%%%%%%
%%%%%%%%%%%%% mod-grav
%%%%%%%%%%%%%%%%%%%%%
 \bibitem{mod-grav} Sean M. Carroll, Vikram Duvvuri, Mark Trodden, Michael S. Turner,
 {\it Is Cosmic Speed-Up Due to New Gravitational Physics?}
 Phys.Rev. D70 (2004) 043528

A.D. Dolgov, M. Kawasaki, {\it Can modified gravity explain
accelerated cosmic expansion}, Phys.Lett. B573 (2003) 1-4

M. E. Soussa, R. P. Woodard, {\it The Force of Gravity from a
Lagrangian containing Inverse Powers of the Ricci Scalar},
 Gen.Rel.Grav. 36 (2004) 855-862

G. Allemandi, A. Borowiec, M. Francaviglia, {\it Accelerated
Cosmological Models in First-Order Non-Linear Gravity},  Phys.Rev.
D70 (2004) 043524; G. Allemandi, A. Borowiec, M. Francaviglia,
{\it Accelerated Cosmological Models in Ricci squared Gravity},
 Phys.Rev. D70 (2004) 103503

J. W. Moffat, {\it Modified Gravitational Theory as an Alternative
to Dark Energy and Dark Matter}, astro-ph/0403266

J. D. Barrow, {\it  Sudden Future Singularities}, gr-qc/0403084
Class.Quant.Grav. 21 (2004) L79-L82.

M.C.B. Abdalla, S. Nojiri, S. D.Odintsov, {\it Consistent modified
gravity: dark energy, acceleration and the absence of cosmic
doomsday}, Class.Quant.Grav. 22 (2005) L35.

\bibitem{Witten-SFT}  E.~Witten,
\textit{Noncommutative geometry and string field theory}, Nucl.
Phys. B268 (1986) 253; E.~Witten, \textit{Interacting field theory
of open superstrings}, Nucl.Phys.~B276 (1986) 291.

\bibitem{Kost-Sam}
V. Kostelecky and S. Samuel, \textit{Spontaneous Breaking of
Lorentz Symmetry in String Theory}, Phys. Rev. D 39 (1989) 683.

\bibitem{West} P.~West, \textit {The Spontaneous Compactification of the Closed Bosonic String},
Phys.Lett. B548 (2002) 92-96.



\bibitem{AMZ-PTY}  I.A.~Aref'eva, P.B.~Medvedev and A.P.~Zubarev,
\textit{Background formalism for superstring field theory},
Phys.Lett.~B240 (1990) 356;
%%CITATION = HEP-TH 0209197;%%

C.R.~Preitschopf, C.B.~Thorn and S.A.~Yost, \textit{Superstring
Field Theory},  Nucl.Phys.~B337 (1990) 363;
%%CITATION = NUPHA,B337,363;%%

I.Ya.~Aref'eva,  P.B.~Medvedev and A.P.~Zubarev, \textit{New
representation for string field solves the consistency problem for
open superstring field}, Nucl.Phys.~B341 (1990) 464.
%%CITATION = NUPHA,B341,464;%%


\bibitem{BSZ} N.Berkovits, A.Sen and  B.Zwiebach,
\textit{Tachyon Condensation in Superstring Field Theory},
Nucl.Phys. B587 (2000) 147-178,  hep-th/0002211.

 \bibitem{ABKM} I.Ya.~Arefeva, D.M.~Belov, A.S.~Koshelev, P.B.~Medvedev,
{\it Tahyon Condensation in the Cubic Superstring Field Theory},
Nucl.Phys B, 638 ( 2002) 3-20, hep-th/0011117;
%%CITATION = HEP-TH 0011117;%%
{\it Gauge Invariance and Tahyon Condensation in the Cubic
Superstring Field Theory}, Nucl.Phys B, 638 (2002) 21-40,
hep-th/0107197.
%%CITATION = HEP-TH 0107197;

\bibitem{0102085}
K.~Ohmori, \textit{A Review on Tachyon Condensation in Open String
Field Theories}, hep-th/0102085.
%%CITATION = HEP-TH 0102085;%%
\bibitem{0111208}
I.Ya.~Aref'eva, D.M.~Belov, A.A.~Giryavets, A.S.~Koshelev,
P.B.~Medvedev, \textit{Noncommutative Field Theories and
(Super)String Field Theories}, hep-th/0111208.
%%CITATION = HEP-TH 0111208;%%
\bibitem{0301094}
W.Taylor, {\it Lectures on D-branes, tachyon condensation and
string field theory}, hep-th/0301094.
%%CITATION = HEP-TH 0301094;%%

\bibitem{DW} DeWolfe O., Freedman D.Z., Gubser S.S., Karch A.,
Modeling the fifth dimension with scalars and gravity,
 Phys.Rev. {\bf D62} (2000) 046008, hep-th/9909134.


\bibitem{AJK}
  I.Ya. Aref'eva, L.V.~Joukovskaya and A.S.~Koshelev,
\textit{Time Evolution in Superstring Field Theory on non-BPS
brane.I. Rolling Tachyon and Energy-Momentum Conservation}, JHEP
0309 (2003) 012;
%%CITATION = HEP-TH 0301137;%%
\\
I.Ya. Aref'eva, {\it Rolling tachyon in NS string field theory}, Fortschr. Phys., 51 (2003) 652;\\
I.Ya.~Aref'eva and L.V.~Joukovskaya, \textit{Rolling Tachyon on
non-BPS brane}, Lectures given at  the II Summer School in Modern
Mathematical Physics, Kopaonik, Serbia, 1-12 Sept. 2002.
%%%%%%%%%%%%%%%%%

%%%%%%%%%%%%%%%%%%%%%%%%
%%%%%%%%%%%%% study of eqs.
\bibitem{MZ} N. Moeller, B. Zwiebach, {\it   Dynamics with
Infinitely Many Time Derivatives and Rolling Tachyons}, JHEP 0210
(2002) 034, hep-th/0207107.
%%CITATION = HEP-TH 0207107;%%
\bibitem{BFOW} L. Brekke, P.G.O. Freund, M. Olson,
E.Witten, Nonarchimedean String Dynamics,
 Nucl.Phys. B302, 365, 1988.


\bibitem{Yar} Ya.I.Volovich, {\it Numerical study of
Nonlinear Equations with Infinite Number of  Derivatives}, J.Phys.
A36 (2003) 8685-8702, math-ph/0301028.
%%CITATION = MATH-PH 0301028;%%

\bibitem{VS-YV}  V.S.Vladimirov and Ya.I.Volovich,
\textit{Nonlinear Dynamics Equation in p-Adic String Theory},
Theor.  Math. Phys.,  138 (2004) 297-309, math-ph/0306018.
%%CITATION = MATH-PH 0306018;%%

%%%%%%%%%%%%%%%%two-fields
\bibitem{OH}
 K. Ohmori,
\textit{Toward Open-Closed String Theoretical Description of
Rolling Tachyon}, Phys.Rev. D69 (2004) 026008.

\bibitem{LY}
L. Joukovskaya and Ya.~Volovich, {\it  Energy Flow from Open to
Closed Strings in a Toy Model of Rolling Tachyon},
math-ph/0308034.
%%CITATION = MATH-TH 0308034;%%
%\bibitem{LYLY}
%L. Joukovskaya and Ya.~Volovich,
%{\it  Energy Dissipation in Open-Closed String Model of Rolling Tachyon},
%(in preparation).


%%%%%%%%%%%%%%% Lumps in VSFT
\bibitem{FGN}
V. Forini, G. Grignani, G. Nardelli, {\textit A new rolling
tachyon solution of cubic string field theory}, JHEP 03 (2005)
079, hep-th/0502151.


\bibitem{LB} L. Bonora, C. Maccaferri, R.J.Scherer Santos, D.D.Tolla, {\it
Exact time-localized solutions in Vacuum String Field Theory},
hep-th/0409063.
%%CITATION = HEP-TH 0409063;%%
\bibitem{HM} H. Hata, S.Moriyama, {\it Boundary and Midpoint Behaviors of Lump
Solutions in Vacuum String Field Theory}, hep-th/0504184.
%%CITATION = HEP-TH 0504184;%%

\end{thebibliography}
\end{document}